\documentclass[a4paper,11pt]{article}
\usepackage[utf8]{inputenc}
\usepackage{amssymb,amsmath}
\usepackage{amsfonts}
\usepackage{setspace}
\usepackage{fullpage}
\usepackage[pdftex]{graphicx}
\usepackage{float}
\usepackage{subfigure}
\usepackage{hyperref}
\usepackage{color}
\usepackage{verbatim}
\usepackage[font={small,it}]{caption}
\usepackage[numbers]{natbib}

\def\lsim{\;\raise0.3ex\hbox{$<$\kern-0.75em\raise-1.1ex\hbox{$\sim$}}\;}
\def\gsim{\;\raise0.3ex\hbox{$>$\kern-0.75em\raise-1.1ex\hbox{$\sim$}}\;}
\newcommand{\beq}{\begin{equation}}
\newcommand{\eeq}{\end{equation}}
\def\ben{\begin{enumerate}}  \def\een{\end{enumerate}}
\def\bit{\begin{itemize}}    \def\eit{\end{itemize}}
\def\beq{\begin{equation}}   \def\eeq{\end{equation}}
\def\bea{\begin{eqnarray}}   \def\eea{\end{eqnarray}}
\def\nn{\nonumber}

\def\k{\kappa}
\def\l{\lambda}

\def\D0{D$0$\hskip -6pt/ }

\begin{document}

\begin{titlepage}

\vspace*{-0.25cm}
\begin{flushright}
arXiv:1403.2088
\end{flushright}

\begin{center}
{ \sffamily \Large \bf
Naturalness of scale-invariant NMSSMs with and without extra matter}
\\[8mm]
Maien Y. ~Binjonaid$^{a,b}$,
\footnote{E-mail: \texttt{mymb1a09@soton.ac.uk, maien@ksu.edu.sa}}
and Stephen F.~King$^{a}$
\footnote{E-mail: \texttt{king@soton.ac.uk}}
\\[3mm]
{\small\it
$^a$ School of Physics and Astronomy, University of Southampton,\\
Southampton, SO17 1BJ, U.K.\\[2mm]
$^b$ Department of Physics and Astronomy, King Saud University,\\
Riyadh 11451, P. O. Box 2455, Saudi Arabia
}\\[1mm]
\end{center}
\vspace*{0.5cm}

\begin{abstract}
\noindent
We present a comparative and systematic study of
the fine tuning in Higgs sectors in three scale-invariant NMSSM models: the first being
the standard $Z_3$-invariant NMSSM; the second is the NMSSM plus
additional matter filling $3(5+\overline{5})$
representations of $SU(5)$ and is called the NMSSM+; while the third model comprises
$4(5+\overline{5})$
and is called the NMSSM++. 
Naively, one would expect the fine tuning in the plus-type models to be 
smaller than that in the NMSSM since the presence of extra matter
relaxes the perturbativity bound on $\lambda$
at the low scale. This, in turn, allows larger tree-level Higgs
mass and smaller loop contribution from the stops.
However we find that LHC limits on the masses of sparticles, 
especially the gluino mass, can play an indirect, but vital, role in
controlling the fine tuning. In particular, working in a semi-constrained framework
at the GUT scale, 
we find that the masses of third generation stops are always larger in the 
plus-type models than in the NMSSM without extra matter. This is 
an RGE effect which cannot be avoided, and as a consequence the fine tuning 
in the NMSSM+ ($\Delta \sim 200$) is significantly larger than in the NMSSM ($\Delta \sim 100$), with
fine tuning in the NMSSM++ ($\Delta \sim 600$) being significantly larger than in the NMSSM+.


\end{abstract}

\end{titlepage}
\newpage

\section{Introduction} \label{sc:intro}

The scalar particle discovered in July 2012 \cite{Aad:2012tfa,Chatrchyan:2012ufa}
is increasingly consistent with a Standard-Model-like Higgs boson \cite{Chatrchyan:2013lba}. 
This may reinforce the Hierarchy problem and the call for new physics at low scales just above the Electroweak
scale \cite{PhysRevD.13.3333,Susskind:1978ms}.
Low scale supersymmetry (SUSY) is perhaps the most well-motivated candidate for such new physics beyond the
Standard Model (SM) since it provides, for e.g., a solution to the Hierarchy problem,
a candidate for Dark Matter and unifies the SM group at the Grand Unification (GUT) scale. 
However low scale SUSY remains elusive 
at the LHC \cite{CMS-PAS-SUS-13-019}. 

The naturalness problem in SM \cite{'tHooft:1979bh} 
is associated with the large ratio between the weak scale ($M_W$) and the Planck scale
($M_P$). If no new physics enters at the weak
scale or the TeV scale, then the Higgs mass has to be fine tuned against
the Planck scale, GUT scale, or any new scale represented by possible
heavy masses (e.g. a heavy right-handed neutrino). This situation
is theoretically unpleasant and the lightness of the Higgs needs to
be explained or maintained without huge fine tuning. Supersymmetry
(SUSY) can resolve this issue by cancelling the quadratic divergence
associated with fundamental scalars. 

Nevertheless, the observed value of the Higgs mass ($m_h \sim 126$ GeV)
already places the minimal supersymmetric extension of the standard
model (the MSSM) in tension with the naturalness requirement since
the tree-level Higgs mass bound $m_h \leq M_Z$ implies that very large
stop masses and mixing is required in order to radiatively increase the Higgs mass to its observed value,
leading to a fine tuning in the permille level (see \cite{Feng:2013pwa} for a general discussion
on Naturalness and SUSY).
Moreover, the lower bound on the gluino mass at the LHC of greater than 1 TeV or so
is exacerbating the situation, since the gluino mass radiatively increases the mass of the stops,
independently of their experimental limit, especially in high scale SUSY models such as
the constrained MSSM (cMSSM) or minimal supergravity (mSUGRA) where the effect of gluino
radiative corrections occurs over a larger energy range (for a general discussion on the Status of SUSY after LHC8
we refer the reader to  \cite{Craig:2013cxa}).

Non-minimal SUSY models, such as the next-to-minimal standard model
(NMSSM) (for a review see \cite{Ellwanger:2009dp}), 
can accommodate a $126$ GeV Higgs boson without requiring
such large stop masses and mixing.
This is because non-minimal models usually introduce additional contributions
to the physical Higgs mass at tree level. In particular, the superpotential 
of the NMSSM contains an F-term interaction ($\lambda \hat{S}\hat{H_u}\hat{H_d}$) 
that couples the up- and down-Higgs doublets with the
SM singlet. This will enhance the Higgs mass with an additional term
proportional to $\lambda$ at tree-level (Equation~\ref{mh} in Section~\ref{sc:models}).
Thus, the fine tuning is expected to be less severe 
than in the MSSM since one does not require large stop loop contributions as is the
case in the MSSM \cite{BasteroGil:2000bw,Ellwanger:2011mu,Kim:2013uxa}. 
However, there is an upper bound on $\lambda$ at the low scale ($\lambda \lesssim 0.7$) 
\cite{King:1995ys} for it to be perturbative to the GUT scale.
This indeed will limit the tree-level enhancement to the Higgs mass in the NMSSM.
Moreover, the increased lower bounds on sparticles
from direct searches at the LHC sets the minimum amount of fine tuning
in the Electroweak sector of all SUSY models, and the NMSSM is no exception.

Adding extra matter to the particle content of the NMSSM has a profound
impact on the phenomenology and predictions of the model. In particular, it allows 
$\lambda$ to be larger at the low scale \cite{Masip:1998jc,Barbieri:2007tu,King:2012is},
while still perturbative to the GUT scale.
Indeed, this can improve the tree-level enhancement to the Higgs mass in comparison
with the NMSSM without extra matter. Conventional wisdom dictates that
increasing $\lambda$ at the low scale, by adding extra matter, reduces the fine
tuning of the model. However, surprisingly, this question has not been fully addressed
in the literature in a $Z_3$-invariant semi-constrained SUGRA framework, as far as we know. 
In this paper, we consider two examples of 
the NMSSM with extra matter, and we find that, although $\lambda$ is increased at the low scale,
neither model leads to a reduction in fine tuning. The two
models are called: the ``NMSSM+'',
which is defined by adding extra matter filling three $(5+\bar{5})$
of $SU(5)$, and the ``NMSSM++'', where four extra $(5+\bar{5})$ matter representations
of $SU(5)$ are added to that present in the NMSSM. 

Although the fine tuning in the ``NMSSM+'' has not been discussed before,
a related model, the ``Peccei-Quinn NMSSM'' with additional three $(5+\bar{5})$ states
of $SU(5)$ has been considered  \cite{Barbieri:2007tu}, where the fine tuning due to the parameter
$A_{\lambda}$ (the trilinear soft SUSY breaking term associated with $\lambda$) 
was discussed. This model is characterised by removing the cubic 
self-coupling term of the singlet superfield
($\frac{\kappa}{3}\hat{S}^3$) from the superpotential. On the other hand, the analysis
in \cite{Hall:2011aa} considered a non-scale invariant version of the Peccei-Quinn NMSSM,
as well as the so-called ``$\lambda$-SUSY'' model, where $\lambda$ is not required
to be perturbative to the GUT scale, but only to $\sim 10$ TeV
to comply with electroweak precision tests.
Further references will be given in Section~\ref{sc:ft}.

In this paper, then, we study and compare the fine tuning in
three $Z_3$-invariant semi-constrained GUT models: the NMSSM,
where we update previous literature, 
and the NMSSM+ and NMSSM++ for the first time. We show that, surprisingly, 
while $\lambda$ assumes larger values in the plus-type models than 
in the NMSSM, hence the tree-level Higgs mass is larger in such models, 
there is an indirect RGE effect,
played by the gluino, that renders the plus-type models more fine tuned than 
the NMSSM. 
\footnote{In particular, we find that, in order to obtain the same physical gluino
mass at the low scale in the three models, the GUT scale boundary condition of
the gluino mass parameter $M_3(M_{\text{GUT}})$ will follow a specific ordering.
Namely, $M_3(M_{\text{GUT}})$ is larger in the NMSSM+, and 
even larger in the NMSSM++, as compared to the NMSSM.} As a consequence of this
unavoidable RGE effect (explained in Section~\ref{sc:1lp}), the mass of the stops
will always be larger than in the NMSSM+, and even 
larger in the NMSSM++, as compared to the NMSSM. Taking into account current 
LHC limits and constraints on the Higgs, 
third generation squarks, and the gluino, the lowest fine tuning 
in the semi-constrained NMSSM, NMSSM+, and NMSSM++ is 
found to be about 100, 200 and 600, respectively, which is a new and unexpected result. 
While the NMSSM and the NMSSM+ 
are less fine tuned than the cMSSM, 
the NMSSM++ is fine tuned to a level comparable to that in the cMSSM.
More importantly, our results show that increasing the perturbativity bound on $\lambda$
by adding extra matter does not reduce the fine tuning. In fact,
it can increase the fine tuning significantly.   

In Section \ref{sc:models}, we give a brief overview of the models is given. Section \ref{sc:1lp} discusses
certain one-loop RGEs and features of each model. In Section \ref{sc:ft},
we discuss the fine tuning measure that is used, and the two-loop RGEs implementations.
Next, we discuss the theoretical framework at the GUT scale, and the ranges of parameter space we are considering 
in each model in Section \ref{sc:frame}.  
Section \ref{sc:res} is where we present our main results.
Finally, we conclude in Section \ref{sc:con}.

\section{The models} \label{sc:models}

Non-minimal models are associated with adding fields not present in
the SM, and/or enlarging the gauge structure. The NMSSM is a well-known
example where the $\mu$ term in the MSSM is omitted, and a
SM-singlet field is introduced. This field acquires VEV near the weak
scale to dynamically generate a $\mu$ effective term. The NMSSM keeps
all the good features of the MSSM, such as unification of gauge couplings,
and radiative Electroweak Symmetry Breaking. It is also known to have
lower fine tuning than the MSSM as mentioned in Section \ref{sc:intro}. However, to avoid
unwanted weak-scale Axion, one introduces a cubic term for the singlet
and the superpotential is invariant under a discrete $Z_3$ symmetry,

\begin{equation} \label{eq:w}
\mathcal{W}_{\text{NMSSM}} = \frac{\kappa}{3}\hat{S}^3 + \lambda \hat{S}\hat{H_1}\hat{H_2} + \mathcal{W}_{\text{MSSM}}(\mu=0)  
\end{equation} 
where $\hat{H_1} = \hat{H_d},$ $\hat{H_2} = \hat{H_u}$ are the down- and up-type Higgs superfields, $\hat{S}$ 
is a SM singlet superfield. $\kappa$ is the cubic coupling of the singlet,
and $\lambda$ is the Higgs singlet-doublet coupling.
$\mathcal{W}_{\text{MSSM}}(\mu=0)$ is the superpotential of the MSSM without a $\mu$ term. 
Note that \ref{eq:w} is invariant under a discrete $Z_3$ symmetry, and once the VEVs are acquired this symmetry 
is broken. The consequence of such breaking will be discussed at the end of this Section. 

The Higgs and the SM singlet superfields will acquire VEVs represented classically as,

\begin{equation}
\langle H_1 \rangle =    \begin{pmatrix} v_1 \\ 0 \end{pmatrix}, \ \ \  \langle H_2 \rangle =  
\begin{pmatrix} 0 \\ v_2 \end{pmatrix},  \langle S \rangle = v_3 , 
\end{equation}

In terms of these VEVs, the scalar Higgs potential reads, 

\begin{equation} \begin{split} V_{\text{NMSSM}} = & m_1^2 v_1^2 + m_2^2 v_2^2 + \lambda^2 v_1^2 v_2^2 
+ 2\mu_{\text{eff}} B_{\text{eff}} v_1 v_2 \\ & + \frac{ \bar{g}^2 }{ 8 } (v_1^2 - v_2^2)^2 + v_3^2(m_S^2 
+ \frac{2}{3} k v_3 A_{\kappa} + \kappa^2v_3^2). \end{split} \end{equation} 

where, $m_j^2 = m_{H_j}^2 + \mu_{\text{eff}}^2$, for $j=1,2$. $\mu_{\text{eff}} = \lambda v_3$ and 
$B_{\text{eff}} = \kappa v_3 + A_{\lambda}$ are effective terms produced as the SM singlet 
acquires its VEV. $A_{\lambda}$ and $A_{\kappa}$ are trilinear soft terms
associated with the couplings $\lambda$ and $\kappa$. $m_S$ is the soft mass of the singlet.
And $\bar{g}^2 = g_1^2 + g_2^2$, where $g_1$ and $g_2$ are the gauge couplings
associated with $U(1)_Y$ and $SU(2)_L$, respectively.  

From the minimisation conditions, $\frac{\partial V}{\partial v_i} = 0$, where the index $i$ runs from 1 to 3,
we obtain three conditions for Electroweak Symmetry Breaking in terms of the mass of the $Z$ boson, $M_Z$, 
and $\sin 2\beta$, where $\tan \beta = \frac{v_2}{v_1}$,
and the soft mass of the SM singlet, $m_S$:

\begin{equation} \frac{M_Z^2}{2} =  \frac{ m_1^2 - \tan^2\beta m_2^2 }{\tan^2\beta - 1}, \label{ewsb1}\end{equation}
\begin{equation} \sin 2 \beta = \frac{2 \mu_{\text{eff}} B_{\text{eff}}  }{m_1^2 + m_2^2 + \lambda^2 v^2}, \label{ewsb2}
\end{equation}
\begin{equation} m_S^2 + \kappa A_{\kappa} v_3 + \kappa^2 v_3^2 \simeq 0 \label{ewsb3} \end{equation}
where, $v^2 = v_1^2 + v_2^2 = (174 \ \text{GeV})^2$. 

Equations~\ref{ewsb1}-~\ref{ewsb2} are similar to those of the MSSM, while Equation~\ref{ewsb3}
is absent in the MSSM since 
it does not contain a SM singlet superfield. In contrast to the MSSM,
the $\mu_{\text{eff}}$ in the NMSSM depends on 
soft parameters as it includes $v_3$, which, in turn, can be written
in terms of $m_S$ and $A_{\kappa}$ by using Equation~\ref{ewsb3}. 

The soft terms, $\{ m_{H_j}, m_S, A_{\kappa},$ and $A_{\lambda} \}$, at the low scale, e.g. 
$M_{\text{SUSY}} \sim \mathcal{O}(1 \text{TeV})$, can be expanded in terms of the fundamental parameters
of the theory that are specified at the GUT scale using the Renormalisation Group Equations (RGEs) 
(this will be briefly discussed in Section \ref{sc:1lp}).
In particular, in the framework of mSUGRA/CNMSSM, all 
scalar masses share a common mass: $m_0$, all gaugions share a common mass: $m_{1/2}$, and all 
trilinear couplings share a common value: $A_0$. This is called universal boundary conditions. One can work 
on a framework where some or all of this universality is relaxed. 

One of the remarkable features of the NMSSM is that it allows for the increase
of the tree-level Higgs physical mass via an additional F-term contribution:

\begin{equation} m_h^2 \leq M_Z^2 \cos^2 2\beta + \lambda^2 v^2 \sin^2 2 \beta, \label{mh}\end{equation} 
therefore, unlike the case in the MSSM, moderate values of $\tan\beta$ ($<$10) are preferred in conjunction
with large values of $\lambda \sim 0.7$. Additionally, loop corrections to the physical Higgs mass, which 
are dominated by the top/stop, need not be as large as in the MSSM. This means that, in the NMSSM, the A-term
can be as small as zero, and the lightest stop can be significantly smaller than in the MSSM (more discussion
can be found in \cite{King:2012is}). Moreover, it is well-known that in the NMSSM, the SM-like Higgs can 
be either the lightest or the next-to-lightest CP-even Higgs states.   

Nevertheless, the NMSSM is also known to have its own issues, namely,
the ``domain wall problem'' that arises as the $Z_3$ symmetry is spontaneously
broken near the Electroweak scale \cite{Vilenkin:1984ib}. This problem, as well as 
the 0.7 bound on $\lambda(M_{\text{SUSY}})$ are the main motivation for studying 
extensions of the NMSSM where extra matter surviving to a scale of a few TeV are present. 
Plus-type models can overcome both issues \cite{Hall:2012mx} and offer 
a link to a more fundamental (F-Theory) framework \cite{Callaghan:2013kaa}. 

In the notation of $SU(5)$ representations,
the two models we are considering and comparing with the NMSSM can be viewed as:

\begin{equation}\text{NMSSM+} \approx \ \text{NMSSM} \ \ + \ 3(5+\bar{5}), \end{equation}
and
\begin{equation}\text{NMSSM++} \approx \ \text{NMSSM} \ \ + \ 4(5+\bar{5}).\end{equation}

\begin{figure}[H]
\begin{minipage}[t][1\totalheight][c]{0.45\columnwidth}%
\includegraphics[scale=0.35]{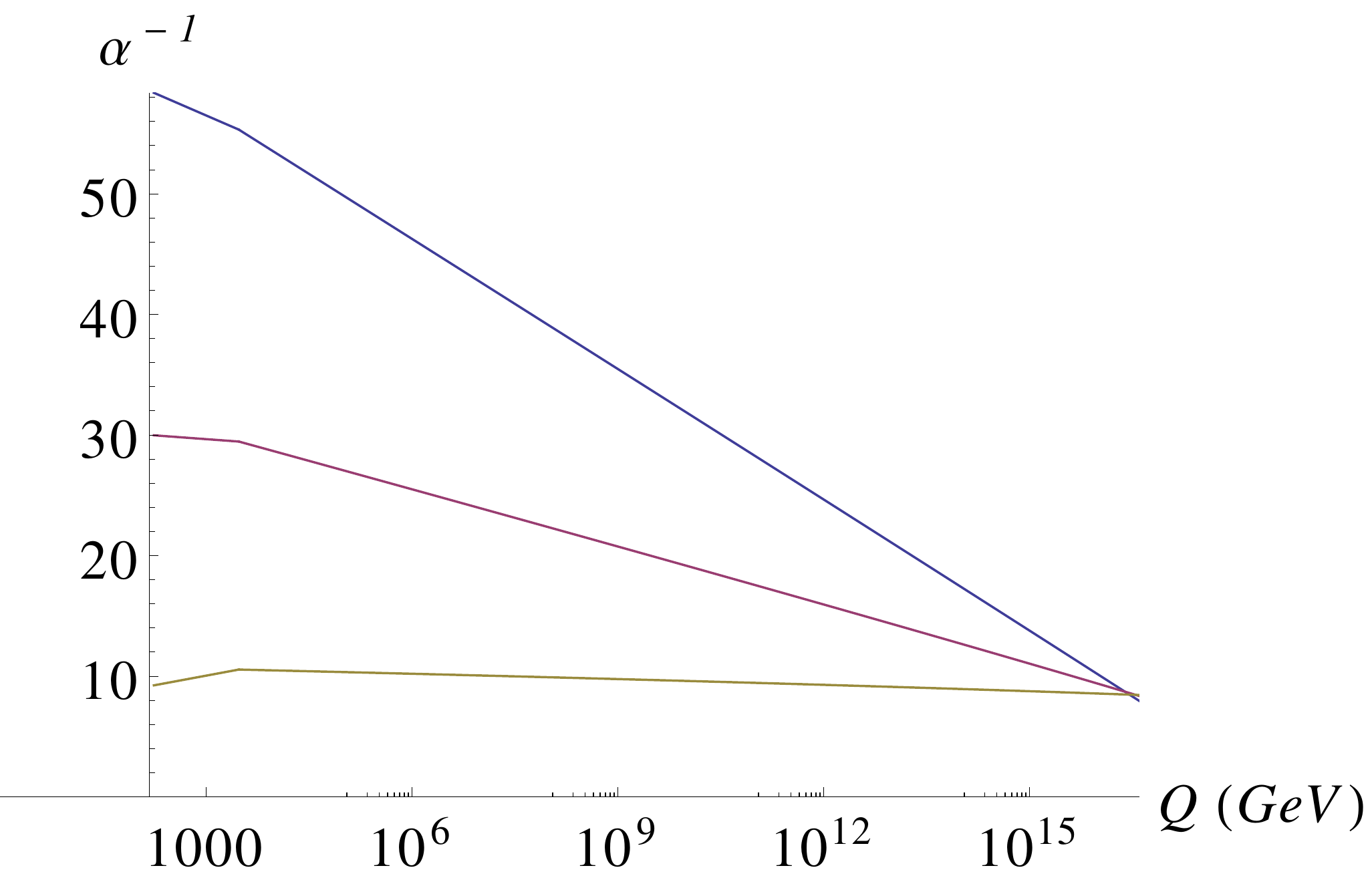}%
\end{minipage}\hfill{}%
\begin{minipage}[t][1\totalheight][c]{0.45\columnwidth}%
\includegraphics[scale=0.35]{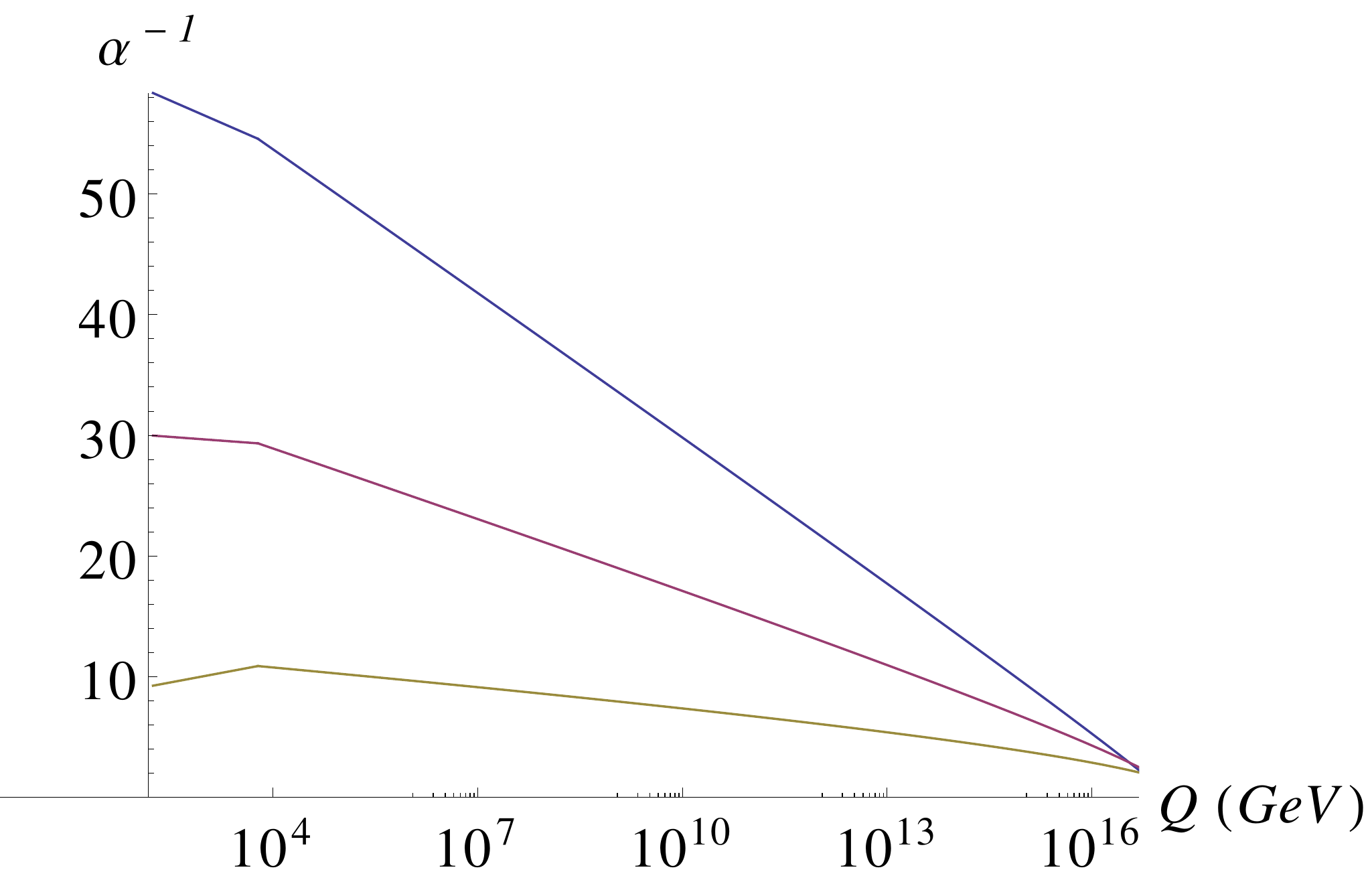}%
\end{minipage}
\caption{Gauge coupling unification to two-loop in the NMSSM+ (Left) and the NMSSM++ (Right). 
The mass scale of the extra matter in the NMSSM+ (NMSSM++) is taken here to be 3 (6) TeV. Below that scale, the NMSSM
without extra matter is assumed.
At $M_t=173.6$ GeV, we set: $g_{1,\text{SM}} =0.35940$, $g_2=0.64754$, $g_3=1.1666$, $h_t=1.01685$, $\tan\beta=5$, $\lambda=0.7$,
and
$\kappa=0.1$.  }
\label{fig:uni}
\end{figure}

From low energy standpoint, Eqs.~\ref{ewsb1}-\ref{mh} hold in the plus-type models
to a good approximation, this is because the extra matter reside in a secluded sector that 
only relates to ordinary NMSSM superfields through gauge interactions, this is a key feature of 
the models we are considering, and as a consequence, the chief effect 
of the presence of the extra matter is the modification of running of the gauge couplings (and gaugino mass running) 
at one-loop, and the running of the rest of the parameters at two-loop. Gauge coupling unification is
approximately achieved at two-loop in both 
plus-type models (Figure~\ref{fig:uni}) since the extra matter form complete representations 
of SU(5). Furthermore, in the NMSSM+ (++), and for a mass scale of the extra matter of 3 (6)
TeV, the unification scale is $M_{\text{GUT}} \sim 2.5 \times 10^{16}$ GeV ($M_{\text{GUT}} \sim 3.6 \times 10^{16}$ GeV), and
the unified coupling is $\alpha_{\text{GUT}}\sim 0.11$ ($\alpha_{\text{GUT}}\sim 0.33$). This can be compared to the NMSSM, where $M_{\text{GUT}} \sim 1.5 \times 10^{16}$ GeV, and $\alpha_{\text{GUT}} \sim 0.04$. Moreover, the implication
of such increase in $M_{\text{GUT}},$ and $\alpha_{\text{GUT}}$ is that the proton lifetime from dimension-6
operators ($\tau_p \propto \frac{M_{\text{GUT}}^4}{\alpha_{\text{GUT}}^2}$) will be roughly, $\tau_p \sim 2.5 \times 10^{34} $ years ($\tau_p \sim 1 \times 10^{34} $ years), in comparison to the NMSSM where $\tau_p \sim 2.1 \times 10^{34} $ years.

In Section~\ref{sc:1lp} we provide a comparison of specific one-loop RGEs and approximate solutions in order to 
establish some crucial differences between the models that will be relevant in subsequent Sections. 

\section{One-loop renormalisation group analysis } \label{sc:1lp}

In this Section we present a one-loop analysis of the three models to illustrate a few key points
that will aid in anticipating and understanding the fine tuning results in Section \ref{sc:res}.
The main arguments will still be valid even though we incorporate two-loop RGEs in our analysis in Section \ref{sc:res}.

\begin{figure}[H]
\begin{minipage}[t][1\totalheight][c]{0.45\columnwidth}%
\includegraphics[scale=0.30]{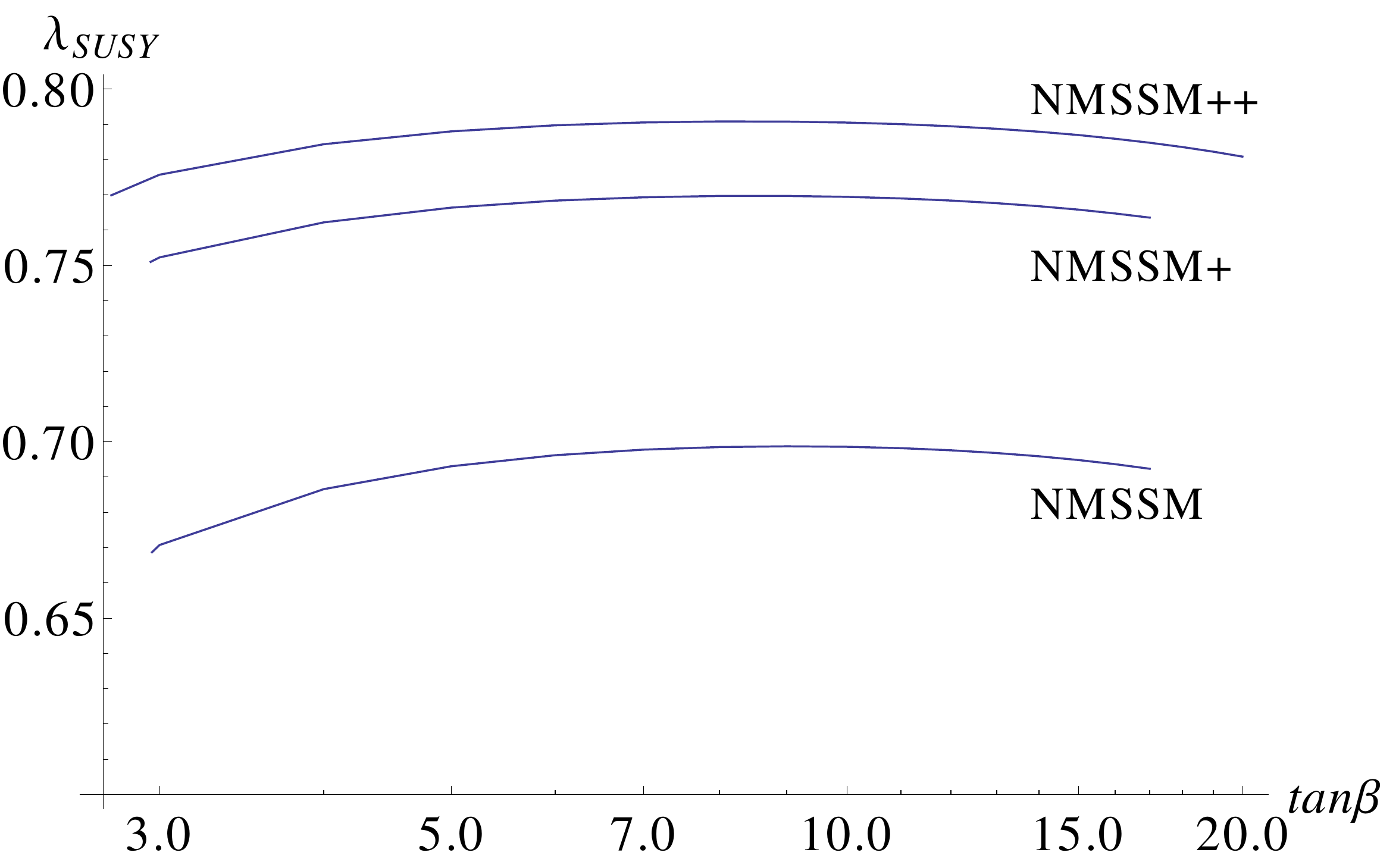}%
\end{minipage}\hfill{}%
\begin{minipage}[t][1\totalheight][c]{0.45\columnwidth}%
\includegraphics[scale=0.30]{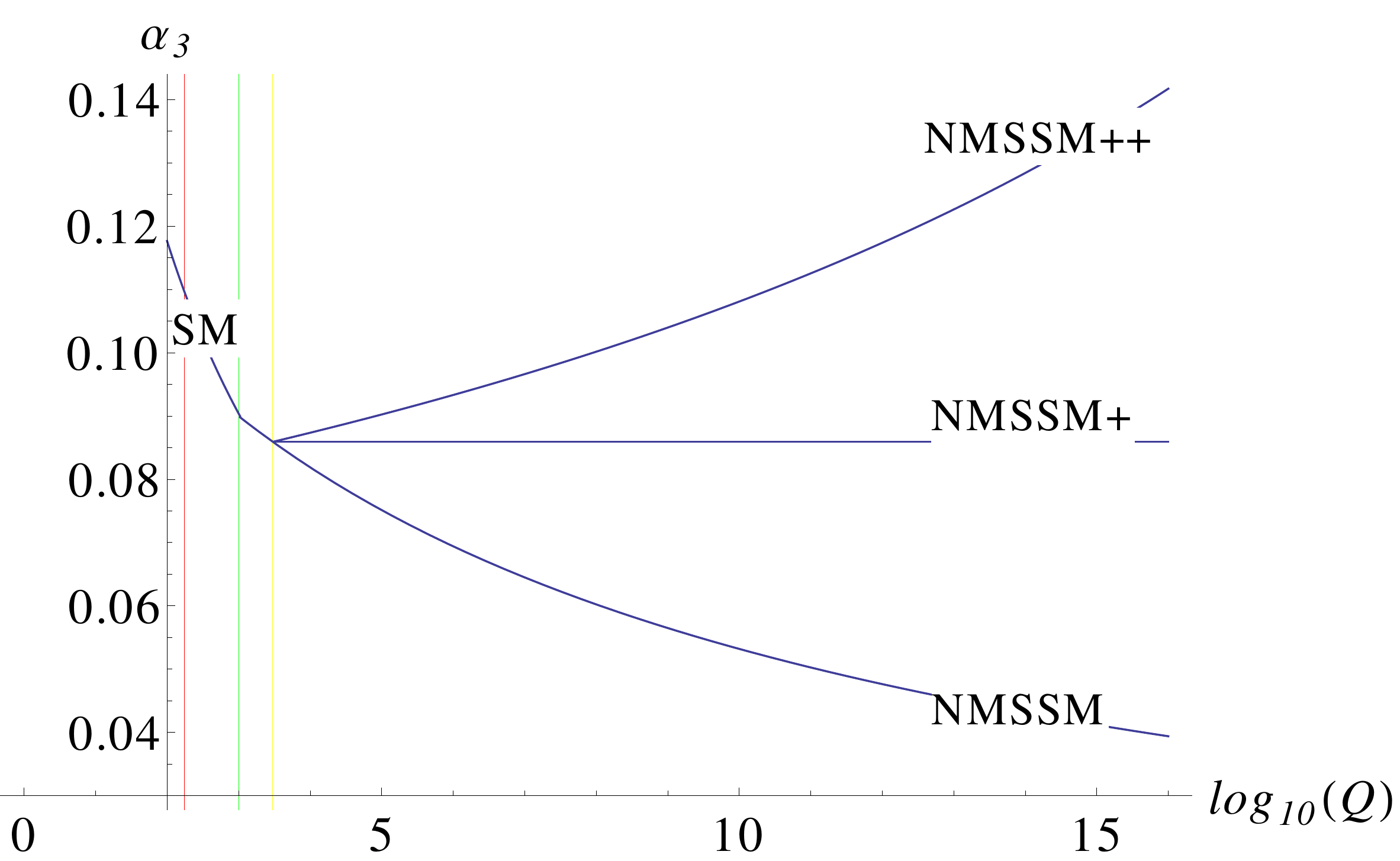}%
\end{minipage}

\caption{Left panel: $\lambda(1 \ \text{TeV})$ as a function of $\tan \beta$ 
in the three different models (for $\kappa(1 \ \text{TeV}) \sim 0.002$). 
Right panel: The one-loop running of the strong coupling $\alpha_3(\equiv \alpha_s)$, where
the running from $M_Z$ (vertical black line), passing through $m_t=173.6$ GeV (vertical red line), to a fixed SUSY scale at 1 TeV (vertical green line) is performed using SM RGE. Next, the running from 1 TeV to the GUT scale is performed using
the NMSSM RGE, and at 3 TeV (vertical yellow line) the NMSSM+ and the NMSSM++ RGEs are used to run
up to the GUT scale.}
\label{fig:lam}
\end{figure}

The addition of extra matter in the plus-type models is motivated
both from the high scale and the low scale model building point of view. In particular,
by examining the effects on the RGEs, one can show that the perturbativity
bound on $\lambda$ at the SUSY scale ($\lambda_{M_{\text{SUSY}}}$) increases in the plus-type
models as shown in the left panel of Figure~\ref{fig:lam}.

The reason behind this increase can be understood by inspecting the
RGEs of the gauge couplings: $g_a$ (where $a=1,2$ and 3, for $U(1)_Y$, $SU(2)_L$, and
$SU(3)_c$ gauge groups, respectively), the top Yukawa coupling:
$h_t$, the doublet-singlet coupling: $\lambda$, in the three models.
At one-loop, the RGEs take the following form:

\begin{equation} 16 \pi^2 \partial_t g_1^2 = \left(11,16,\frac{53}{3} \right) g_1^4 \label{eq:beta1}\end{equation}
\begin{equation} 16 \pi^2 \partial_t g_2^2 = \left(1,4 ,5 \right)g_2^4 \label{eq:beta2} \end{equation}
\begin{equation} 16 \pi^2 \partial_t g_3^2 = \left(-3,0,1\right)g_3^4  \label{eq:beta3} \end{equation}
\begin{equation} 16 \pi^2 \partial_t h_t^2 = h_t^2 \left(6h_t^2 + h_b^2 + \lambda^2 -\frac{13}{9}g_1^2 - 3g_2^2 -\frac{16}{3}g_3^2  \right) \end{equation}
\begin{equation} 16 \pi^2 \partial_t \lambda^2 = \lambda^2 \left(3h_t^2 + 3h_b^2 + 4\lambda^2 -g_1^2 -3g_2^2 \right),  \end{equation}
where, $\partial_t \equiv \frac{\partial}{\partial \ln Q^2}$, and $Q$ is the renormalisation scale. 
The coefficients between parentheses in Equations~\ref{eq:beta1}-\ref{eq:beta3} belong to the NMSSM, the
NMSSM+ and the NMSSM++, respectively. And $g_1$ is SM normalized (as opposed to the GUT normalization that introduces
a factor of $\sqrt{\frac{3}{5}}$, i.e. ${g_1^2}_{,\text{SM}} = \frac{3}{5}{g_1^2}_{,\text{GUT}}$). 
The magnitudes and the signs
of these $\beta$-function coefficients lead to larger $g_3$
and smaller $h_t$, at the GUT scale, in the NMSSM+ compared to the NMSSM, 
and similarly larger values of both couplings
in the NMSSM++ compared to the NMSSM+, at the GUT scale.
This allows larger $\lambda$ at the low scale (e.g. 1 TeV) while keeping its
perturbativity to the GUT scale. The advantage of having
a larger low-scale $\lambda$ is that it allows for a larger tree-level
Higgs mass in \ref{mh}. Moreover, since the top/stop Yukawa coupling depends on
$\sin\beta$ as follows, 
\begin{equation} h_t (Q) = \frac{m_t(Q)}{v \sin\beta}, \end{equation}
it is possible to achieve smaller $\tan\beta$ in the plus-type
models.

Moreover, it is instructive to examine the running of $\alpha_3 = \frac{g_3^2}{4\pi}$
(which runs similar to the gluino mass parameter $M_3$). This is
shown in the right panel of Figure~\ref{fig:lam}. Note that, in order
to reach the same point at the low scale, say $\alpha_3(1 \ \text{TeV})$, in three models, the starting point at the GUT scale 
(i.e. the boundary condition: $\alpha_{3,\text{\tiny{GUT}}} \equiv \alpha_3(M_{\text{GUT}}$) is significantly different.
In particular,
\begin{equation}
 \alpha_{3,\text{\tiny{GUT}}}^{\text{\tiny{NMSSM++}}} > \alpha_{3,\text{\tiny{GUT}}}^{\text{\tiny{NMSSM+}}} 
> \alpha_{3,\text{\tiny{GUT}}}^{\text{\tiny{NMSSM}}}.
\end{equation}
This effect will play a profound role
in shaping the fine tuning (as we show in Section \ref{sc:res}) since we expect a similar 
behaviour in the running of the gluino mass parameter $M_3$. And although we 
use two-loop RGEs to obtain our fine tuning results in Section \ref{sc:res}, the argument 
is still valid, namely that, in order to reach the same physical gluino mass
at the low scale in the three models, the GUT scale boundary condition $M_{3,\text{\tiny{GUT}}}\equiv M_3(M_{\text{GUT}})$
will follow the ordering:
\begin{equation} \label{eq:M3}
 M_{3,\text{\tiny{GUT}}}^{\text{\tiny{NMSSM++}}} > M_{3,\text{\tiny{GUT}}}^{\text{\tiny{NMSSM+}}} > 
M_{3,\text{\tiny{GUT}}}^{\text{\tiny{NMSSM}}}.
\end{equation}

The physical gluino mass at the low scale can be approximately related to the input
parameter $m_{1/2}$, which is a universal gaugino mass at the GUT
scale, as follows: $m_{\tilde{g}} \approx f m_{1/2}$, where the coefficient
$f$ is model-dependent. We will present these values in Section \ref{sc:res}. 
Next, we consider the implication of the ordering in \ref{eq:M3}.
The gluino affects the running of the squarks at one-loop
in the following fashion, 
\begin{equation}
\frac{\partial m_{\tilde{Q_3}}^2 }{\partial t} = -\frac{3 \alpha_3}{8 \pi} M_3^2 + f(m_{\text{scalars}}^2,A^2,g_a^2, \dots), \label{eq:run}
\end{equation}
where we are only showing the gluino mass term explicitly.
It is well-known \cite{Kane:1998im,Hardy:2013ywa,Arvanitaki:2013yja} 
that the gluino mass parameter, if large enough at the GUT scale,
can dominate the running of the scalars. It is also well-known that coloured scalars 
run from the GUT scale to the low scale in such a way that the
running masses increase. Any negative term in the RGE will enhance this 
increase in the running mass at the low scale, and indeed the
gluino mass term in Equation~\ref{eq:run} is negative, thus the larger
the boundary condition ($M_{3,\text{\tiny{GUT}}}$) the larger the scalar mass will be at the low scale. 

We wish to point out that, in the MSSM, obtaining a physical Higgs mass of 126 GeV requires very
large stops or large stop mixing (large A-term), hence, it is requiring 
a 126 GeV Higgs that is causing the fine tuning (in addition to direct limits on sparticles). Whereas 
in the NMSSM, the stops do not need to be as large as in the MSSM, but the limits from direct searches, 
especially on the stops will play a crucial role in determining the fine tuning. However, 
in the plus-type models we are considering, we expect that the
stops in the NMSSM+ will always be larger than in the NMSSM, 
and they will always be larger in the NMSSM++ than in the NMSSM+.
This is a result of the rather larger values of the $M_3(M_{\text{GUT}})$,
or $m_{1/2}$, that one has to start with at the GUT scale in order to achieve a gluino mass larger
than 1.2 TeV at the low scale, as indicated in Equation~\ref{eq:M3}. Therefore, we expect that the gluino 
is the main source of fine tuning in the plus-type models,
and we verify that in Section \ref{sc:res}.  

Next, we present approximate solutions of the one-loop RGE of the parameter $m_{H_u}$ in the three models.
This is for $\tan\beta = 2, \kappa(M_{\text{SUSY}})=0.002,$ and $M_{\text{SUSY}} = 1$ TeV,
and we expand $m_{H_u}(M_{\text{SUSY}})$ in terms of universal GUT parameters: $m_{1/2}$, $m_0$ and $A_0$,
\begin{enumerate}
\item NMSSM $\left(\lambda(M_{\text{SUSY}}) = 0.6\right)$:
\begin{equation}\label{mhu1} -m_{H_u}^2(M_{\text{SUSY}}) \approx 3 m_{1/2}^2 + 0.8 m_0^2 + 0.07 A_0^2 - 0.09 m_{1/2} A_0 \end{equation}
\item NMSSM+ $\left(\lambda(M_{\text{SUSY}}) = 0.72\right)$:
\begin{equation}\label{mhu2} -m_{H_u}^2(M_{\text{SUSY}}) \approx 2.04 m_{1/2}^2 + 0.74 m_0^2 + 0.09 A_0^2 - 0.18 m_{1/2} A_0 \end{equation}
\item NMSSM++ $\left(\lambda(M_{\text{SUSY}}) = 0.75\right)$:
\begin{equation}\label{mhu3} -m_{H_u}^2(M_{\text{SUSY}}) \approx 1.78 m_{1/2}^2 + 0.71 m_0^2 + 0.1 A_0^2 - 0.3 m_{1/2} A_0. \end{equation}
\end{enumerate}

While it is clear from Equations~\ref{mhu1}-~\ref{mhu3} that the sensitivity of
$m_{H_u}^2(M_{\text{SUSY}})$ to $m_{1/2}^2$ is reduced by
adding extra matter, it is important to notice the
ordering in Equation~\ref{eq:M3}. Clearly, the more matter
included, the larger the required $m_{1/2}$ in order to produce the
desired physical $m_{\tilde{g}}$, hence the larger the stops. We quantify this to two-loop and study the associated 
fine tuning in the following Sections.

\section{Fine tuning and two-loop implementations} \label{sc:ft}

\subsection{Fine tuning} \label{ssc:ft}
To quantify fine tuning at each point in the parameter space, one can measure 
the fractional sensitivity of an observable, namely the mass of the Z boson, $M_Z$ to 
fractional variations in the fundamental GUT parameters, $a = \{m_{1/2}, m_0,  m_S, m_{H_u},m_{H_d}$, 
$A,  A_\lambda, A_\kappa,$ $\lambda, \kappa, h_t \}$ (\cite{Ellis:1986yg} and \cite{Barbieri:1987fn}),

\begin{equation} 
\Delta_a = \left| \frac{\partial \log M_Z}{\partial \log a}, \right|
\label{eqn:ft}
\end{equation}  
where $\Delta^{-1} \times 100\%$ represents the percentage to which 
a parameter is fine tuned. 

This measure is usually called the Barbieri-Giudice measure, and 
it has been extensively used in the literature (see for e.g. \cite{Kowalska:2014hza, Kaminska:2014wia,  
Kaminska:2013mya,Boehm:2013qva,Lalak:2013bqa,Fichet:2012sn,Gherghetta:2012gb,
Agashe:2012zq}, and \cite{Athron:2013ipa} and references therein). 
Note that some authors prefer to use $M_Z^2$ instead of $M_Z$ and/or $a^2$ instead of $a$. All different choices 
can be related to each other by the inclusion of an appropriate factor. This global sensitivity of Equation~\ref{eqn:ft}, 
alternative measures, and Bayesian approaches has been briefly discussed in \cite{Athron:2013ipa}.

Moreover, the measure (Equation~\ref{eqn:ft}) is already implemented in the Fortran code NMSPEC \cite{Ellwanger:2006rn} 
that we use, which is part of the package 
NMSSMTools 4.1.2. In this package, the fine tuning is calculated in two steps: first, the tuning with respect 
to SUSY scale parameters \beq \label{eq:msusy} m_{H_{u,d}}(M_{\text{SUSY}}), 
m_S(M_{\text{SUSY}}), A_{\lambda, \kappa}(M_{\text{SUSY}}), \lambda(M_{\text{SUSY}}), 
\kappa(M_{\text{SUSY}}), h_t(M_{\text{SUSY}})  \eeq is calculated using Equation~\ref{eqn:ft} with 
the parameter $a$ being a SUSY scale parameter in \ref{eq:msusy}. Second, the results are linked to GUT scale 
parameters using the RGEs, hence determining 
the fine tuning with respect to the GUT scale parameters. The procedure is discussed in details 
in \cite{Ellwanger:2011mu}. This method is equivalent to deriving a fine tuning ``master formula'' 
for the NMSSM, as in \cite{BasteroGil:2000bw}.

\subsection{Two-loop implementations} \label{ssc:2lp}

We modify the tool for both the
NMSSM+ and the NMSSM++ cases by adding the relevant two-loop RGEs (presented in Appendix~\ref{A}) to enable 
calculating the mass spectrum of each model and study the fine tuning.

\begin{figure}[H] 
\begin{center}
\begin{minipage}[t][1\totalheight][c]{0.45\columnwidth}%
\includegraphics[scale=0.15]{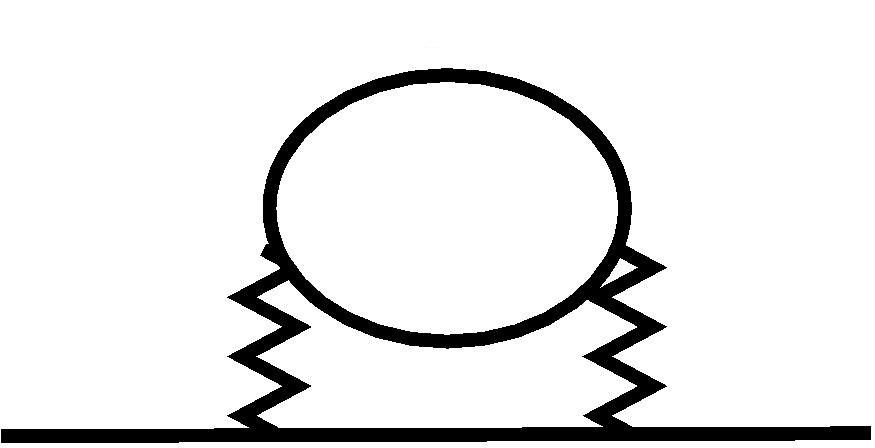}%
\end{minipage}
\end{center}
\caption{A schematic two-loop diagram illustrating how extra matter loops
can modify $\mathcal{O}(g_a^4)$ terms of the running scalars, gauginos, trilinears, and Yukawas 
(see Appendix~\ref{A}) }
\label{fig:2lp1}
\end{figure}

One can start from the two-loop RGEs of the NMSSM, and then modify them for the NMSSM+ and the NMSSM++ cases.
For example, the extra fields, which are charged under the SM gauge group, will change the coefficients 
of $\mathcal{O}(g_a^4)$ terms of the beta functions since they can run in the loop as depicted in Figure~\ref{fig:2lp1}

The relevant terms to be modified are calculated using the results 
in \cite{Martin:1993zk}, and the full set of RGEs are cross-checked using the package SARAH \cite{Staub:2010jh}.  
Furthermore, we take an effective theory approach whereby the extra matter are integrated out via 
a step-function change of the beta functions at a scale
\beq Q_{\text{SUSY}} = \sqrt{\frac{2m_{Q_1}^2 + m_{U_1}^2 + m_{D_1}^2}{4}} , \eeq as defined in NMSSMTools
to be the scale of the first and second generations squarks. $m_{Q_1}^2,m_{U_1}^2,$ and $m_{D_1}^2$
are scalar squared masses of the first generation squarks. In the parameter spaces scanned in Section~\ref{sc:res},
it ranges between $1.1$-$4.1$ TeV in the NMSSM,
$1.9$-$6.4$ TeV in the NMSSM+, and $3.5$-$7.5$ TeV in the NMSSM++.
 
Moreover, for convenience, we assume a degenerate mass scale 
for the extra matter, and we set it to be $Q_{\text{SUSY}}$. Consequently, the RGEs of the full theory, i.e. the NMSSM+(++), 
are used between the GUT scale and the scale $Q_{\text{SUSY}}$, whereas the RGEs of the effective theory, 
i.e. the NMSSM, are used below that. As the RGEs descend from $Q_{\text{SUSY}}$, NMSSMTools includes leading 
logarithmic threshold corrections to the gauge and Yukawa couplings from the relevant 
superpartners. However, since the mass scale of some of the squarks (and all of the extra matter) 
is of order $Q_{\text{SUSY}}$, such states do not contribute to the threshold corrections,
as pointed out in \cite{Ellwanger:2009dp}.

Nevertheless, the extra matter sector is in fact a secluded sector since no Yukawa couplings 
are shared with the NMSSM superfields. The extra matter will obtain fermionic 
and scalar masses in the secluded sector by mechanisms that are irrelevant to 
the weak scale (further details in \cite{Hall:2012mx}).  
As such, we have not calculated the precise mass spectrum of this secluded sector. 
Additionally, in our set-up, the contributions
from the running masses of the extra matter to the NMSSM scalar masses can be safely ignored
at one- and two-loop \footnote{
In $\xi$, $\xi'$, $\sigma_1, \sigma_2$ and $\sigma_3$ in Equation~\ref{app:a3} in Appendix~\ref{A}}
since they are highly suppressed and only introduce a relative error smaller than $1\%$. 

Finally, it worth mentioning that the ordering in Equation~\ref{eq:M3} will remain valid at the two-loop order, 
therefore the situation will always be such that 
for a given physical gluino mass, say 1.2 TeV, the NMSSM+ will require $M_3(M_{\text{GUT}})$ to be 
larger than that in the NMSSM, and hence the stops in the NMSSM+ will be larger 
than the stops in the NMSSM. 
Similarly, the NMSSM++ will require $M_3(M_{\text{GUT}})$ to be larger than that in the 
NMSSM+, which means the stops will be larger in the former than in the latter. 
We verify this and study the implication on the fine tuning in Section~\ref{sc:res}. 
 
\section{Framework and parameter space} \label{sc:frame}
\subsection{Framework} \label{ssc:frame}
We choose to work in a semi-constrained framework where the gaugino
masses are universal at the GUT scale, i.e. $M_1(M_{\text{GUT}}) =
M_2(M_{\text{GUT}}) = M_3(M_{\text{GUT}}) = m_{1/2}$, where $M_1,M_2$ are Bino and Wino 
mass parameters. One the other hand, we allow $m_S$,
$m_{H_u}$ and $m_{H_d}$ to differ from the rest of the scalars that have a common
mass $m_0$ at the GUT scale. However, since we use $\mu_{\text{eff}}$ as an input,
NMSSMTools will output the allowed values for those parameters at
the GUT scale. In addition, the trilinears $A_{\lambda}$ and $A_{\kappa}$
can take different values, at the GUT scale, from the universal trilinear 
$A_0 = A_t(M_{\text{GUT}})=A_b(M_{\text{GUT}})=A_{\tau}(M_{\text{GUT}})$, where the indices $t,b,\tau$ denote 
the top, bottom, and $\tau$ squarks. 

Moreover, it is crucial to note that choosing non-universal gauginos at the GUT scale, i.e. $M_1(M_{\text{GUT}})$ $\neq
M_2(M_{\text{GUT}}) \neq M_3(M_{\text{GUT}})$ might be desirable~\footnote{One possible situation where 
abandoning this universality is desirable is to have $M_1(M_{\text{GUT}}) \neq m_{1/2}$ to satisfy 
dark matter constraints as discussed in \cite{Ellwanger:2011mu}}. However, we do not make 
this assumption here since it has no impact on the fine tuning comparison for the three models.
In particular, as
we show in Section~\ref{sc:res}, $M_3(M_{\text{GUT}})$ controls the fine tuning 
in the plus-type models, while the other two parameters ($M_1$ and $M_2$) have little or no impact.
Hence, we find it simpler to assume universality in our analysis. Finally, we do
not include constraints from dark matter in our analysis (although we check that regions of low fine
tuning are not excluded by an upper bound of $\Omega h^2 < 0.13$ as calculated by the package micrOMEGAs \cite{Belanger:2013oya} that is embedded in NMSSMTools), and we are not addressing the issue of the anomalous magnetic of the muon.

\subsection{Parameter space} \label{ssc:scan}
We have focused on the parameter space where $\lambda$, at the low scale, can be as large as possible,
while $\tan \beta$ can be as small as possible in the three models,
this is subject to constraints from perturbativity, successful Electroweak symmetry breaking, 
and experimental limits, all of which are taken into account in NMSSMTools~\footnote{A full list of constraints can be found in the official website of NMSSMTools:\url{http://www.th.u-psud.fr/NMHDECAY/nmssmtools.html}} (including: LEP bounds on Higgs searches and invisible Z decays,
constraints on new physics from $b\rightarrow s \gamma$, $B_s \rightarrow \mu^+ \mu^-$, and $B \rightarrow \tau \nu_{\tau}$,
all to within 2$\sigma$). The mass of the SM-like Higgs is required to be $m_h = 125.7 \pm 3$ GeV to account for uncertainties. If the SM-like Higgs is the second-to-lightest Higgs in the NMSSM, then
NMSSMTools will ensure that the lightest Higgs satisfies LEP constraints. Furthermore, NMSSMTools ensures that the couplings
and signals of the SM-like Higgs comply with LHC results as studied in \cite{Belanger:2013xza}. Additionally, we require that $m_{\tilde{t}_1} > 700$ GeV,
and $m_{\tilde{g}} > 1.2$ TeV \cite{ATLAS-CONF-2013-062}. Removing the constraint on $m_{\tilde{t}_1}$
from our analysis does not negate our main finding that the NMSSM is less fine tuned than the NMSSM+, and the NMSSM+ is less fine tuned than the NMSSM++. Additionally, it is difficult to relax this constraint since this will depend on certain mass relations (e.g. between $m_{\tilde{t}_1}$ and $m_{\tilde{\chi}^0_1}$), which we are not analysing here {\footnote{It is important to mention that, while the lightest stop plays a role in the determination of the fine tuning, the heavy stop also
plays a role, as well as the trilinear coupling $A_t$, and the soft Higgs mass $m_{H_u}$. For instance,
a very light $m_{\tilde{t}_1}$ can be obtained if $A_t$ is quite large. However, this will lead to
a rather large $m_{\tilde{t}_2}$, which in turn will contribute to the fine tuning via the radiative
corrections to the Higgs potential.}} 
 
We use the simple random sampling method provided by NMSSMTools. However, in order to test the effect of increasing $\lambda$ by adding extra matter on the fine tuning, we choose a representative range of the parameters $\lambda$, $\tan\beta$, and $\mu$ that leads to an enhancement to the tree-level Higgs mass, and to a reduction of the tuning in $M_Z$. Our strategy is to scan small patches of 
the parameter space, with narrow ranges of $m_0$, $m_{1/2}$, and $A$ in order to find solutions where
the fine tuning is expected to be small. With this in mind, we scan up to $6\times 10^7$ points in this region of the parameter space in each model. Next, points that violate the constraints mentioned previously are removed. Finally, we divide the data into two sets,
the first set is where the lightest Higgs is SM-like, and the second set is where the second-to-lightest Higgs
is SM-like. The scanned range of parameters is,

{\centering
$0 < m_0 <\left(2,4,7\right)$ TeV \\
$0 < m_{1/2} <\left(2,4,7\right)$ TeV \\
$-3.5 < A < 7 $ TeV \\
$-3.5 < A_{\lambda} < 3.5 $ TeV \\
$-3.5 < A_{\kappa} < 3.5 $ TeV \\
$100 < \mu_{\text{eff}} < 400$ GeV \\
$0 < \tan \beta < 5$ \\
$0.5 < \lambda < 1$ \\
$10^{-4} < \kappa < 0.6$ \\}
\noindent
where the numbers between parentheses in the first two lines correspond to
the range in the NMSSM, the NMSSM+, and the NMSSM++, respectively. 
In all models, the fine tuning plots range from 0 to 2000 --we stop at $\Delta = 2000$ for convenience-- 
using the same colour scheme. This enables direct comparison between the parameter spaces of the three models.

\section{Results} \label{sc:res}

In this Section, we present the results for the fine tuning in the parameter spaces of the three models. 
For each model, we have divided the parameter space into two cases, the first (Case 1) is where the SM-like
Higgs is the lightest CP-even Higgs, whereas the second (Case 2) is where the SM-like Higgs is the 
next-to-lightest CP-even Higgs. The reason for this is that the detailed phenomenology of the two
cases can be different (e.g. see \cite{Belanger:2012tt} and \cite{King:2012is}).
    
\subsection{NMSSM } \label{ssc:nmssm}

As stated in Section~\ref{sc:intro}, the NMSSM is well-known to be less fine tuned than the most studied 
supersymmetric model that is the MSSM. Given the current LHC limits on the Higgs couplings, on 
the mass of naturalness-related superpartners, such as the stops and the gluino, the 
results in this section serve as an update to the status of the fine tuning in the NMSSM within 
the range of parameter space specified in Section~\ref{sc:frame}.  

\subsubsection{Case 1: $m_{h_1}$ is SM-like.} \label{sssc:case1}

\begin{figure}[H]
\begin{minipage}[t][1\totalheight][c]{0.45\columnwidth}%
\includegraphics[scale=0.3]{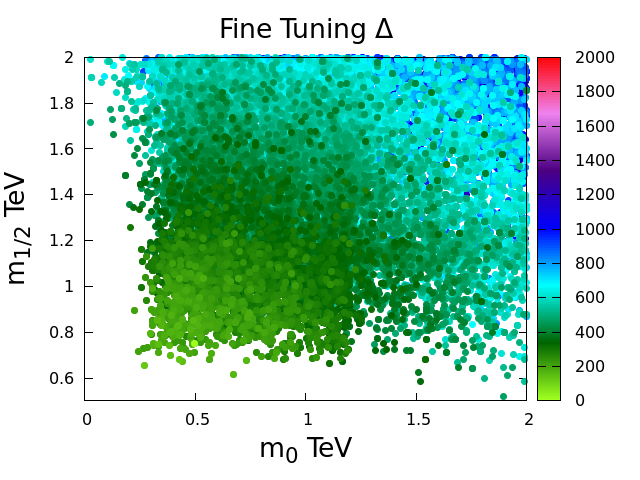}%
\end{minipage}\hfill{}%
\begin{minipage}[t][1\totalheight][c]{0.45\columnwidth}%
\includegraphics[scale=0.3]{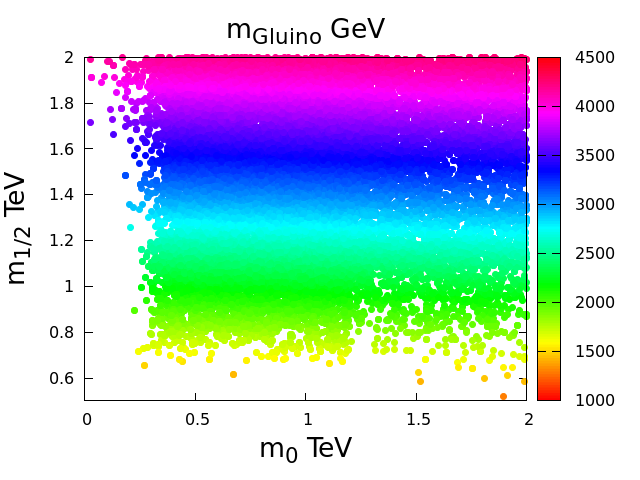}%
\end{minipage}
\caption{Left panel shows the fine tuning while the right panel shows the gluino mass, both
in the $m_0-m_{1/2}$ plane in the NMSSM when $m_{h_1}$ is SM-like.}
\label{fig:nmssm1}
\end{figure}

Figure~\ref{fig:nmssm1} (Left) shows the fine tuning, $\text{max}(\Delta_a)$ or simply $\Delta$, 
represented by colours in the $m_0-m_{1/2}$ plane,
which ranges from 0 to 2 TeV. In this range of parameter space we find that $\Delta \ll 2000$. 
The lowest fine tuning was found to be $\Delta \sim 120$ 
for $m_{h_1} = 124$ GeV, $m_{\tilde{g}} = 1.4$ TeV, $m_{\tilde{t}_1} = 750$ GeV.
Furthermore, the fine tuning forms contours in this plane and the band of contours associated with $120 < \Delta < 300$
corresponds to values of $m_0$ and $m_{1/2}$ that range from 0.3 - 1 TeV and 0.6 - 1.2 TeV, respectively.
As $m_{1/2}$ becomes smaller and approaches 0.5 TeV, $m_0$ becomes larger and approaches 1.9 TeV, thus 
increasing the fine tuning up to $\sim 500$. On the other hand, as $m_0$ becomes 
smaller and approaches zero, $m_{1/2}$ rises to around 1.7 TeV. Consequently, the fine tuning rises above 600. 
Additionally, regions where $m_0$ and $m_{1/2}$ are both above 1.3 TeV are associated with $\Delta > 500$. 
In particular, at the top-right corner where both $m_0$ and $m_{1/2}$ are of $\mathcal{O}(2 \ \text{TeV})$, $\Delta \sim 1000$. 
It is worth-noting that this parameter space is in fact multidimensional since all fundamental parameters 
assume different values at each point.  

A number of observables is significantly linked with fine tuning,
this includes: $m_{h_1}, m_{\tilde{g}},$ and $m_{\tilde{t}_{1,2}}$. In the NMSSM, 
the lowest fine tuning ranges from 100 to 200 for a
Higgs mass between 123 and 127 GeV. 

The gluino mass (Right panel of Figure~\ref{fig:nmssm1}) in this parameter space form 
plateaus specified by the value of the parameter $m_{1/2}$. In particular, $m_{\tilde{g}}$ ranges between $\sim 1.4$ TeV and 
2 TeV for values of $m_{1/2}$ between 0.5 TeV and 0.8 TeV, and increases gradually with $m_{1/2}$ 
to reach values of order 4.5 TeV as $m_{1/2}$ reaches 2 TeV. In fact, by examining the data one finds 
that $m_{\tilde{g}} \sim 3 m_{1/2}$ in the NMSSM. This will remain true for Case 2 in~\ref{sssc:case2}.  

\begin{figure}[H]
\begin{minipage}[t][1\totalheight][c]{0.45\columnwidth}%
\includegraphics[scale=0.3]{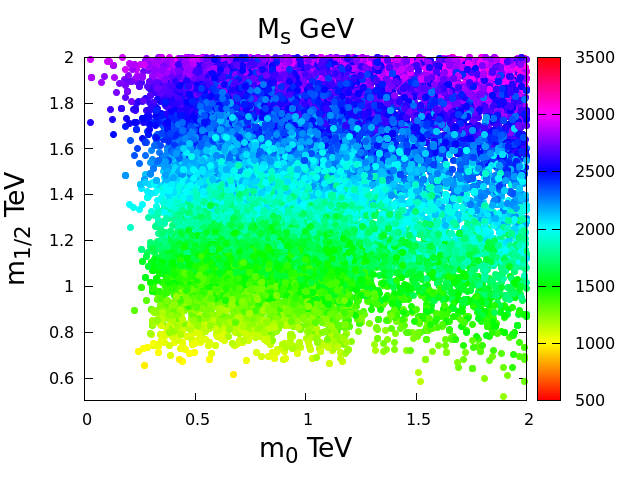}%
\end{minipage}\hfill{}%
\begin{minipage}[t][1\totalheight][c]{0.45\columnwidth}%
\includegraphics[scale=0.3]{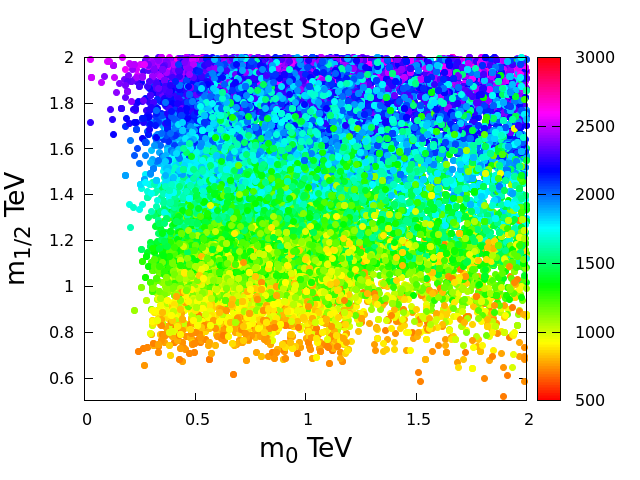}%
\end{minipage}
\caption{Left panel shows the RMS stop mass, while the right panel shows the 
lightest stop mass, both in the $m_0-m_{1/2}$ plane in the NMSSM when $m_{h_1}$ is SM-like. }
\label{fig:nmssm2}
\end{figure}

For convenience, we define the root-mean-square (RMS) stop mass, which we will frequently use,
\beq M_S = \sqrt{ \frac{m_{\tilde{t}_1}^2 + m_{\tilde{t}_2}^2 }{2} } \label{eq:ms} \eeq
and we plot it in the $m_0-m_{1/2}$ plane. Since we require the lowest
mass for the lightest stop to be larger than 700 GeV, $M_S$ can tell
us if there is much separation between $m_{\tilde{t}_1}$ and $m_{\tilde{t}_2}$.
Our aim is to search for points where both masses are close to 700
GeV or with the minimum separation since such points are associated
with low fine tuning. From the left panel in Figure~\ref{fig:nmssm2}, 
$M_S$ starts at nearly 900 GeV, and increases steadily until
reaching 3.4 TeV with increasing $m_{1/2}$. However, it increases very slowly in respond to an increase in $m_0$, 
particularly in this range of parameter space. 
  
In the right panel of Figure~\ref{fig:nmssm2},
the distribution of the lightest stop mass $m_{\tilde{t}_1}$ shows that it ranges from 700
GeV to $\sim 2.7$ TeV. Also, it grows steadily with increasing $m_{1/2}$. 

\begin{figure}[H]
\begin{center}
\begin{minipage}[t][1\totalheight][c]{0.45\columnwidth}%
\includegraphics[scale=0.3]{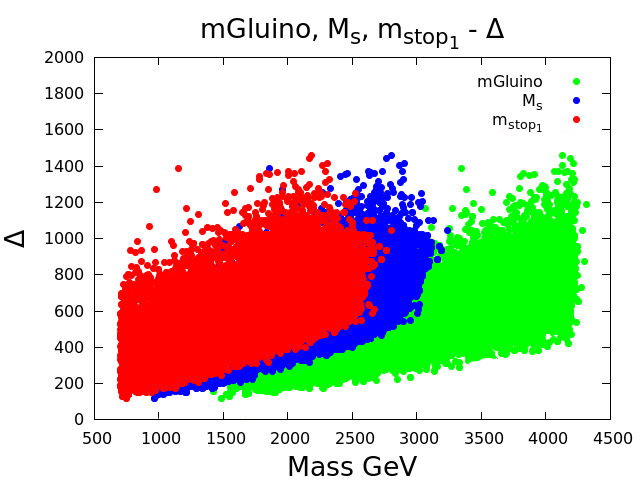}%
\end{minipage}
\end{center}
\caption{Fine tuning as a function of $m_{\tilde{g}}, M_S,$ and $m_{\text{stop}_1}$
in the NMSSM when $m_{h_1}$ is SM-like.}
\label{fig:nmssm3}
\end{figure}

Moreover, figure~\ref{fig:nmssm3}
presents the fine tuning against $m_{\tilde{g}}, M_S$ and $m_{\tilde{t}_1}$.
Notice how the data points of each parameter correlate with the
lowest fine tuning. In particular, $M_S, m_{\tilde{g}},$ and $m_{\tilde{t}_1}$
increase from 900 GeV to 3 TeV, 1.2 TeV to 4.3 TeV and 700 GeV to
2.5 TeV, fine tuning increases from 100 to 600, 400 and 600, respectively.
Clearly, the stop plays the dominant role in determining the
fine tuning. Thus, the gluino can become as large as 4.3 TeV without 
impacting the fine tuning as much as the stops.

The impact of increasing $m_{1/2}$ (and $m_{\tilde{g}}$) on the stops, represented
by $M_S$, will turn to be more significant in the
plus-type models. In the NMSSM, having a gluino mass of 1.2 TeV
does not require $m_{1/2}$ to be larger than $\sim 600$ GeV -recall
that $m_{1/2}$ determines, along with other parameters, the value
of the stops via its RGE effect- and the stops can be as light as
700 GeV. Varying both $m_{1/2}$ and $m_{\tilde{g}}$ from 600 GeV
to 2 TeV and 1.2 TeV to 4.3 TeV, corresponds to $M_S$ in the range 900 GeV-3.4 TeV. 
Therefore, one can escape the LHC limit on the gluino mass without dragging the stops to too heavy masses. 

\subsubsection{Case 2: $m_{h_2}$ is SM-like.} \label{sssc:case2}

\begin{figure}[H]
\begin{minipage}[t][1\totalheight][c]{0.45\columnwidth}%
\includegraphics[scale=0.3]{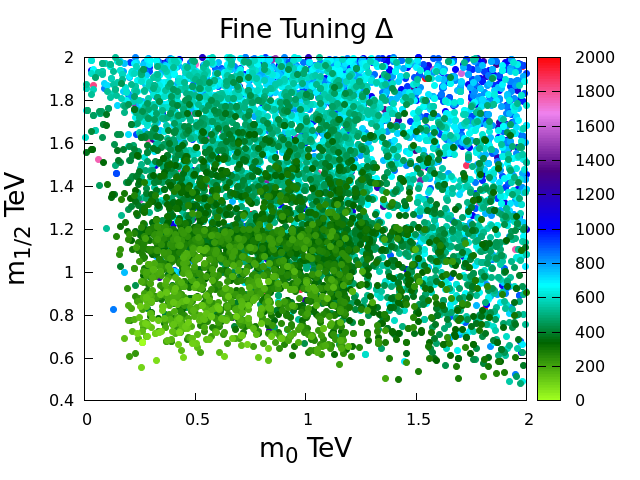}%
\end{minipage}\hfill{}%
\begin{minipage}[t][1\totalheight][c]{0.45\columnwidth}%
\includegraphics[scale=0.3]{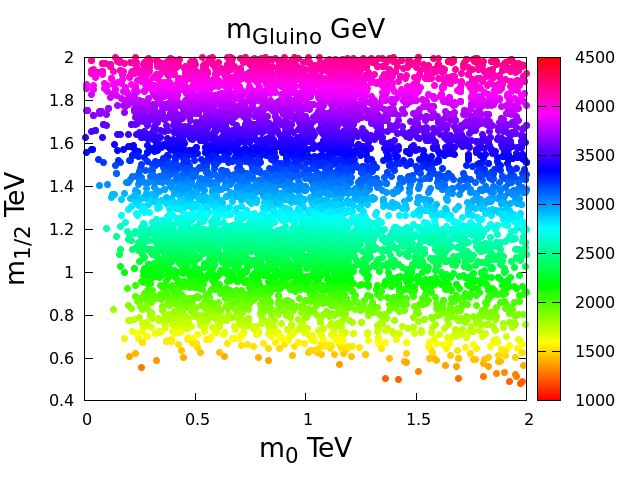}%
\end{minipage}
\caption{Left panel shows the fine tuning while the right panel shows the gluino mass, both
in the $m_0-m_{1/2}$ plane in the NMSSM when $m_{h_2}$ is SM-like.}
\label{fig:nmssm6}
\end{figure}

Here we present the results of fine tuning in the Higgs sector of
the NMSSM where the next to lightest Higgs, $m_{h_2}$, is SM-like. 
First, we note that the fine tuning, Figure~\ref{fig:nmssm6} (left panel),
is roughly similar to Case 1 in~\ref{sssc:case1}. However, the lowest fine tuning here was found to be
$\Delta \sim 71$ 
for: $m_{h_2} = 127$ GeV, $m_{\tilde{g}} = 1.34$ TeV, $m_{\tilde{t}_1} = 700$ GeV,
which is slightly smaller than Case 1 in~\ref{sssc:case1} because $m_{\tilde{t}_1}$ is slightly smaller.
Moreover, in this parameter space, 
we find valid points where $m_{1/2}$ 
can assume lower values than found in the previous case, 
and more points occupying regions where $m_0 = 0$. Those points
at $m_0 \sim 0$ are particularly associated with $A_{\lambda}(M_{\text{GUT}}) > 1$ TeV,
and $100 \leq \mu_{\text{eff}} \leq 260$.

As for the gluino mass (Right panel of Figure~\ref{fig:nmssm6}), it ranges from 
$\sim 1.3$ TeV to 4.4 TeV, and it correlates to $m_{1/2}$ as expected from 
the approximate relation $m_{\tilde{g}} \sim 3 m_{1/2}$. Notice that 
increasing $m_0$ can have a small effect on raising $m_{\tilde{g}}$. This 
is a loop effect related to quark/squark corrections to the physical gluino mass.

\begin{figure}[H]
\begin{minipage}[t][1\totalheight][c]{0.45\columnwidth}%
\includegraphics[scale=0.3]{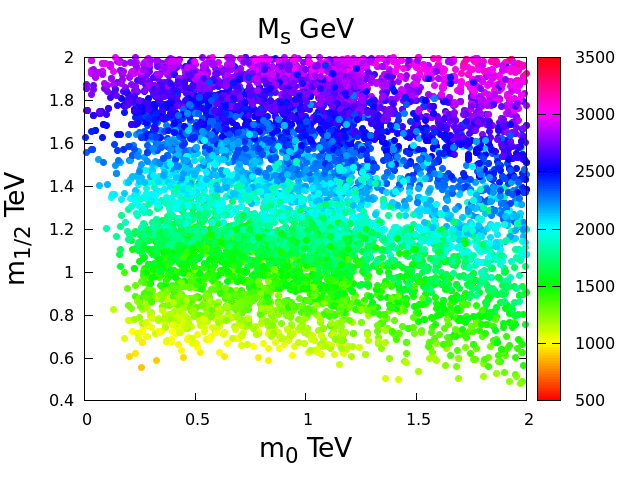}%
\end{minipage}\hfill{}%
\begin{minipage}[t][1\totalheight][c]{0.45\columnwidth}%
\includegraphics[scale=0.3]{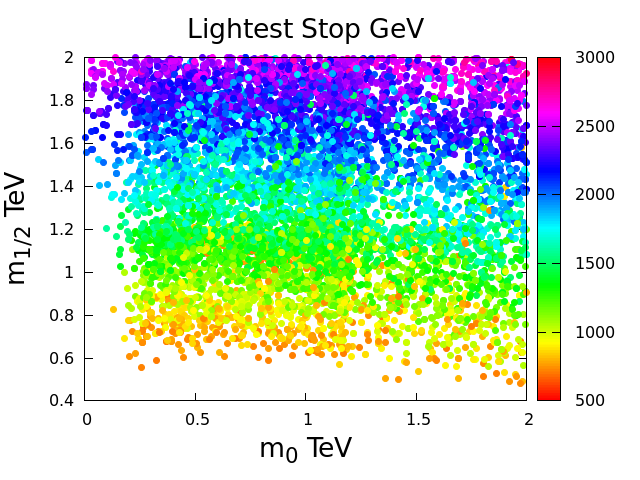}%
\end{minipage}
\caption{Left panel shows the RMS stop mass, while the right panel shows the lightest 
stop mass, both in the $m_0-m_{1/2}$ plane in the NMSSM when $m_{h_2}$ is SM-like.}
\label{fig:nmssm7}
\end{figure}

The lowest fine tuning forms a plateau, of order 100, as one increases
$m_{h_2}$ from 123 GeV to 127 GeV.
Next, Figure~\ref{fig:nmssm7} (left panel) shows how
the parameter $M_S$ varies in the $m_0-m_{1/2}$ plane; it ranges
from 900 GeV to 3.4 TeV, while the right panel shows that the lightest
stop varies between 700 GeV and 3 TeV.

\begin{figure}[H]
\begin{center}
\begin{minipage}[t][1\totalheight][c]{0.45\columnwidth}%
\includegraphics[scale=0.3]{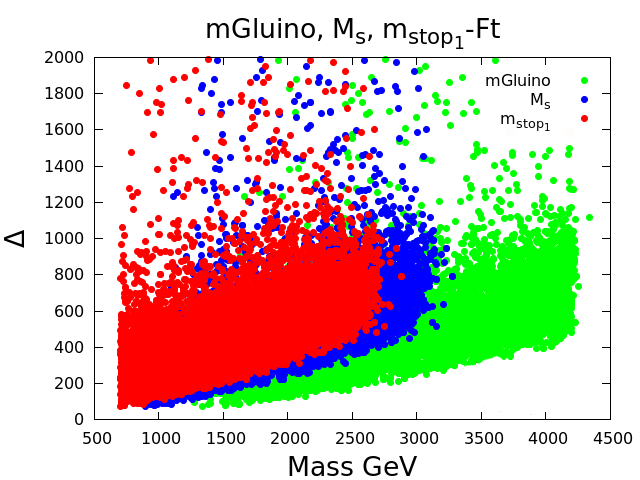}%
\end{minipage}
\end{center}
\caption{Fine tuning as a function of $m_{\tilde{g}}, M_S,$ and $m_{\text{stop}_1}$
in the NMSSM when $m_{h_2}$ is SM-like.}
\label{fig:nmssm8}
\end{figure}

Moreover, figure~\ref{fig:nmssm8} shows 
that increasing $M_S$, $m_{\tilde{g}}$, and $m_{\tilde{t}_1}$ from 900 GeV to 3.3 TeV, 1.2 TeV to 4.3 TeV, 
and 700 GeV to 2.8 TeV results in a rise in the lowest fine tuning from 71 to roughly 450
in the three cases. Therefore, it is still clear that the stops are in control of the fine tuning, 
whereas the gluino mass can assume a value as large as 4.3 TeV without worsening the situation.   

While this parameter space contains the lowest fine tuned point in all our study,
it is still of $\mathcal{O}(100)$, and the parameter space is not as rich as the 
previous one.

\subsection{NMSSM+} \label{ssc:nmssmp}

As discussed in Section \ref{sc:1lp}, the gaugino mass parameter
$m_{1/2}$ has to be larger in the NMSSM+ than
in the NMSSM in order to produce the same physical gluino mass at
the low scale. Moreover, the RG running of scalars depends strongly
on the parameter $M_3$, which is equal to $m_{1/2}$ at the GUT scale.
Therefore, larger $m_{1/2}$, as required by the gluino, means larger
stops, as dictated by the RGEs. Thus, we expect the fine tuning to
be larger in the NMSSM+ than in the NMSSM because the stops are heavier.
The following results show for the first time the fine tuning in the 
Higgs sector of the NMSSM+ with a $Z_3$-invariant superpotential.

\subsubsection{Case 1: $m_{h_1}$ is SM-like.} \label{sssc:case1p}

The parameter space of the NMSSM+ is richer than that of the NMSSM.
In particular, it is easier to obtain a Higgs mass near 126 GeV since
both $\lambda$ and the stops are larger in the NMSSM+ than in the
NMSSM. 

\begin{figure}[H]
\begin{minipage}[t][1\totalheight][c]{0.45\columnwidth}%
\includegraphics[scale=0.3]{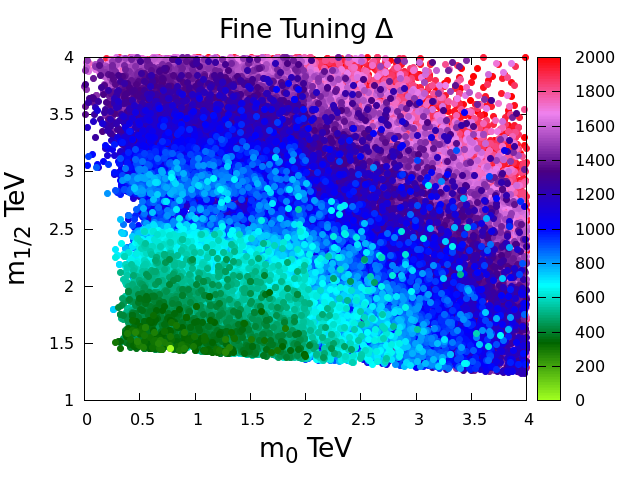}%
\end{minipage}\hfill{}%
\begin{minipage}[t][1\totalheight][c]{0.45\columnwidth}%
\includegraphics[scale=0.3]{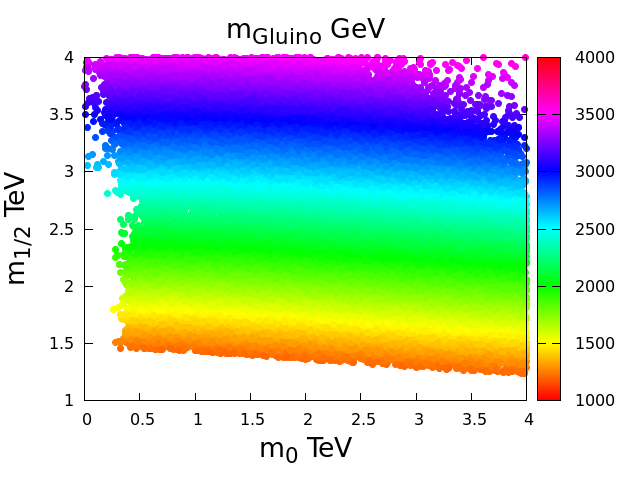}%
\end{minipage}
\caption{Left panel shows the fine tuning while the right panel shows the gluino mass, both
in the $m_0-m_{1/2}$ plane in the NMSSM+ when $m_{h_1}$ is SM-like.}
\label{fig:nmssmp1}
\end{figure}

Figure~\ref{fig:nmssmp1} shows the fine tuning (left panel) distribution in the $m_0-m_{1/2}$
plane, which ranges from 0 to 4 TeV each. Only a relatively small area, located at 
the bottom-left corner, corresponds to fine tuning between 200 and 400. As both 
$m_0$ and $m_{1/2}$ grow larger than 2 TeV, the fine tuning steadily exceeds 400 
reaching values up to 2000. The fine tuning contours show how the fine tuning 
is more sensitive to changes in $m_{1/2}$ than in $m_0$. However, as $m_0$ becomes 
larger than 3.5 TeV, the fine tuning rapidly increases. Regions where $m_0 = 0$ 
are not associated with low fine tuning since they correspond to large values of $m_{1/2}$.
The lowest fine tuning is $\Delta \sim 205$ 
for: $m_{h_1} = 126$ GeV, $m_{\tilde{g}} = 1.2$ TeV, $m_{\tilde{t}_1} = 727$ GeV.

Moreover, notice how the physical
gluino mass in the right panel of Figure~\ref{fig:nmssmp1}
is associated with larger values of $m_{1/2}$ than in the NMSSM
(Figure~\ref{fig:nmssm1}) as explained in Section \ref{sc:1lp}. Particularly, one requires 
$ 1.3 \ \text{TeV} \ < m_{1/2} < 1.5 \ \text{TeV} $ to achieve $ m_{\tilde{g}} \approx 1.2$ TeV.
And the approximate relation between the two parameters is: $m_{\tilde{g}} \sim 0.85 m_{1/2}$.  

\begin{figure}[H]
\begin{minipage}[t][1\totalheight][c]{0.45\columnwidth}%
\includegraphics[scale=0.3]{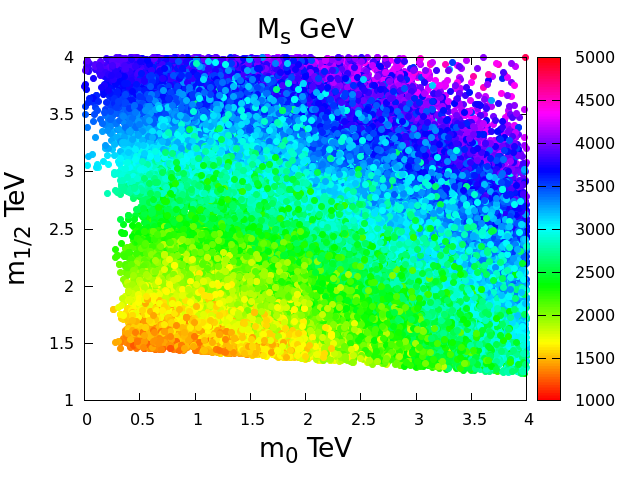}%
\end{minipage}\hfill{}%
\begin{minipage}[t][1\totalheight][c]{0.45\columnwidth}%
\includegraphics[scale=0.3]{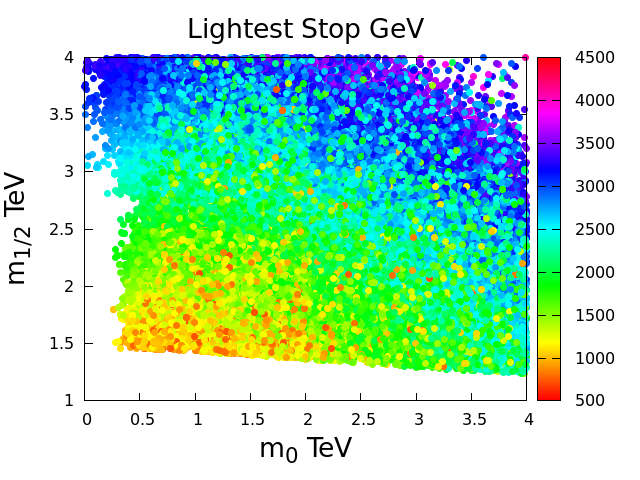}%
\end{minipage}
\caption{Left panel shows the RMS stop mass, while the right panel shows
the lightest stop mass, both in the $m_0-m_{1/2}$ plane in the NMSSM+ when $m_{h_1}$ is SM-like.}
\label{fig:nmssmp2}
\end{figure}

As a result of having a rather large $m_{1/2}$, the smallest value of the
parameter $M_S$ is now around 1.2 TeV (Figure~\ref{fig:nmssmp2}, left panel). 
One can also see that it is not possible to access smaller values of 
$M_S$ because either $m_{1/2}$ or $m_0$ will become exceedingly large. Recall 
that the scalar masses are controlled by both parameters as explained in Section \ref{sc:1lp}. 
The right panel of Figure~\ref{fig:nmssmp2} presents the mass distribution of lightest stop.
It can be as small as 700 GeV and as large as 4 TeV. It worth recalling 
that not only $m_0$ and $m_{1/2}$ determine $m_{\tilde{t}_1}$ and $m_{\tilde{t}_2}$,
but also $A_0$. Large values of 
$A_0$ can lead to large splitting between the lightest and heaviest stops. Therefore,
the data points in Figure~\ref{fig:nmssmp2} (right panel) where small $m_{\tilde{t}_1}$
corresponds to large $M_S$ (left panel), hence large $m_{\tilde{t}_2}$, are associated
with large $A_0$. 

Both $m_{\tilde{t}_1}$ and $m_{\tilde{t}_2}$ contribute to the fine tuning. 
Hence, it is necessary to look at the parameter $M_S$ to understand the fine tuning results.
As stated previously, the larger $M_S$ becomes, the more the fine tuning required. 

As was the case in the NMSSM, varying the Higgs mass between 123 GeV and 127 GeV 
has a little impact on the lowest fine tuning in the NMSSM+. However, the lowest fine tuning
here forms a plateau around $\Delta \sim 200$. 

\begin{figure}[H]
\begin{center}
\begin{minipage}[t][1\totalheight][c]{0.45\columnwidth}%
\includegraphics[scale=0.3]{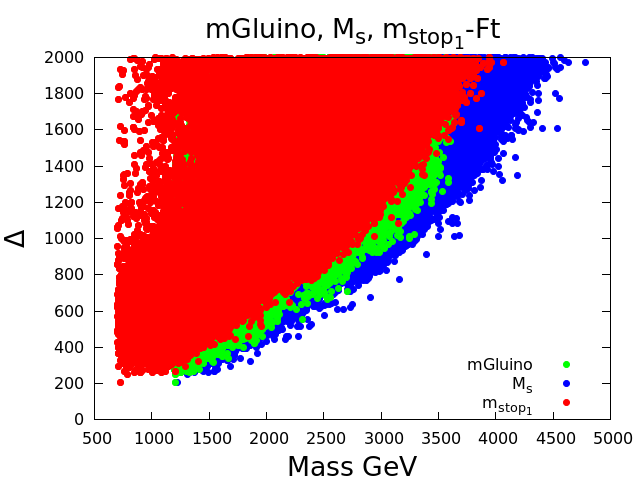}%
\end{minipage}
\end{center}
\caption{Fine tuning as a function of $m_{\tilde{g}}, M_S,$ and $m_{\text{stop}_1}$
in the NMSSM+ when $m_{h_1}$ is SM-like.}
\label{fig:nmssmp3}
\end{figure}

Moreover, $M_S$, $m_{\tilde{g}}$, and $m_{\tilde{t}_1}$ 
(Figure~\ref{fig:nmssmp3}) cause
the lowest fine tuning to increase from 200 to roughly 2000 as they rise
from 1.2 TeV to 4.2 TeV, 1.2 TeV to 3.7 TeV, and 700 GeV to 4 TeV, respectively. 
The important feature that distinguishes the NMSSM+ from the NMSSM is the steady to sharp increase 
in the lowest fine tuning associated with increasing the gluino mass (c.f. Figure~\ref{fig:nmssm3}). 
The lightest stop can now become  
more massive than the gluino and still leads to the same amount of the lowest fine tuning, in contrast to the 
situation in the NMSSM. Clearly, the gluino here is a major factor in determining 
the fine tuning since it requires a large $m_{1/2}$, which in turn affects the 
running of the stops, making them larger in comparison to the NMSSM. 

\subsubsection{Case 2: $m_{h_2}$ is SM-like.} \label{sssc:case2p}

\begin{figure}[H]
\begin{minipage}[t][1\totalheight][c]{0.45\columnwidth}%
\includegraphics[scale=0.3]{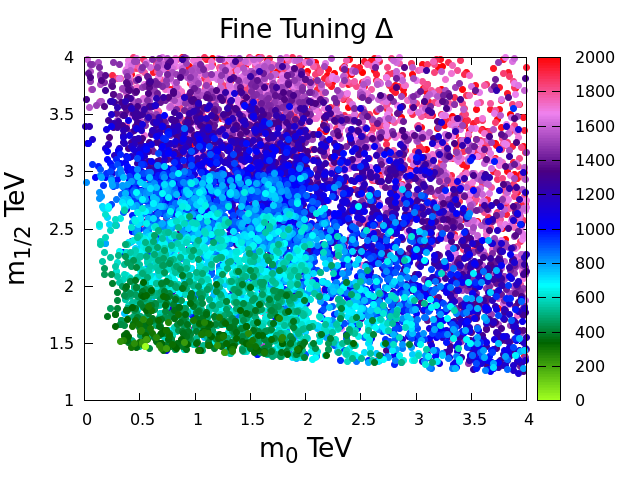}%
\end{minipage}\hfill{}%
\begin{minipage}[t][1\totalheight][c]{0.45\columnwidth}%
\includegraphics[scale=0.3]{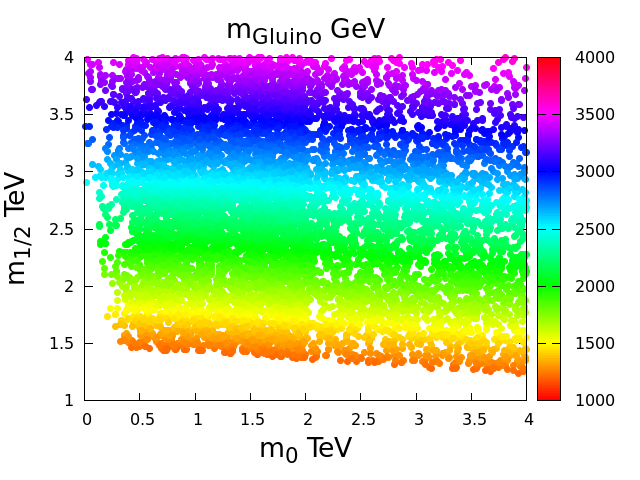}%
\end{minipage}
\caption{Left panel shows the fine tuning while the right panel shows the gluino mass, both
in the $m_0-m_{1/2}$ plane in the NMSSM+ when $m_{h_2}$ is SM-like.}
\label{fig:nmssmp6}
\end{figure}

Here we examine the parameter space of the NMSSM+ where $m_{h_2}$
is SM-like. Figure~\ref{fig:nmssmp6} shows the fine tuning, which starts from
about 188 and reaches 2000, in the $m_0-m_{1/2}$ plane. The features of the fine tuning are similar 
to those found in Case 1 in~\ref{sssc:case1p}.
However, more points can reach the $m_0 = 0$ region in this parameter space.
Points close to $m_0 \sim 100$ GeV, and between $2 \ \text{TeV} <m_{1/2}<3$ TeV are
particularly associated with $A_0 > 2400$ TeV.  
Moreover, the lowest fine tuning that was found is $\Delta \sim 188$ 
for: $m_{h_2} = 126.5$ GeV, $m_{\tilde{g}} = 1.2$ TeV, $m_{\tilde{t}_1} = 793$ GeV. 

\begin{figure}[H]
\begin{minipage}[t][1\totalheight][c]{0.45\columnwidth}%
\includegraphics[scale=0.3]{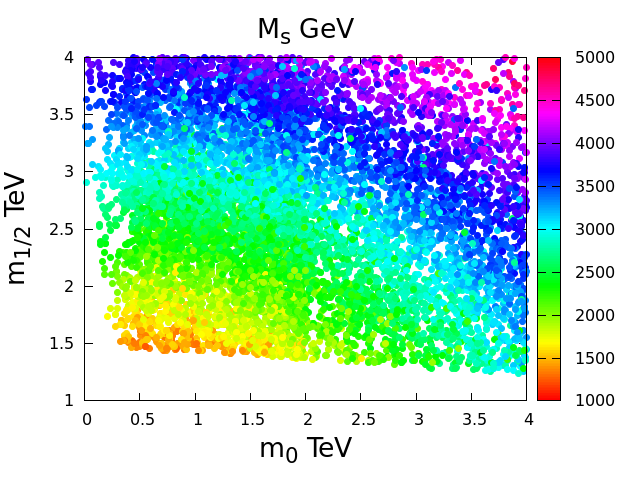}%
\end{minipage}\hfill{}%
\begin{minipage}[t][1\totalheight][c]{0.45\columnwidth}%
\includegraphics[scale=0.3]{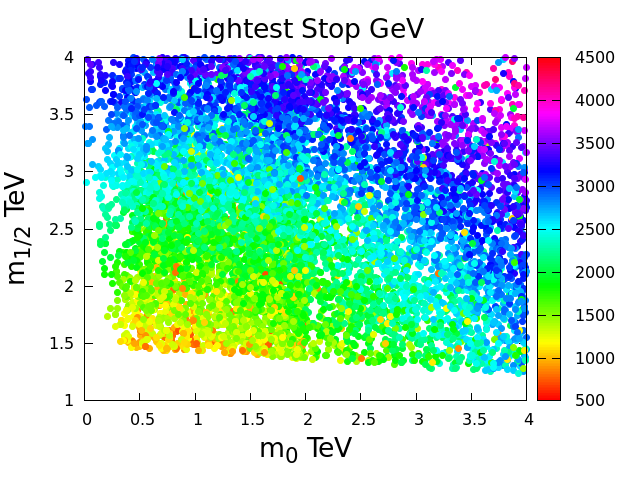}%
\end{minipage}
\caption{Left panel shows the RMS stop mass, while the right panel shows the lightest stop mass, both in the $m_0-m_{1/2}$ plane in the NMSSM+ when $m_{h_2}$ is SM-like.}
\label{fig:nmssmp7}
\end{figure}

The gluino mass (Figure~\ref{fig:nmssmp6}, right panel) also ranges from 1.2 to 3.7 TeV, 
and shares the same features as in Case 1. Again, it correlates to $m_{1/2}$ as: $m_{\tilde{g}}=0.85m_{1/2}$. 
Next, the average stop mass, $M_S$, starts from 1.2 TeV and approaches 5 TeV (Figure~\ref{fig:nmssmp7}, left panel).
On the other hand, the lightest stop mass (right panel) takes values between 700 GeV and 4.2 TeV.  

\begin{figure}[H]
\begin{center}
\begin{minipage}[t][1\totalheight][c]{0.45\columnwidth}%
\includegraphics[scale=0.3]{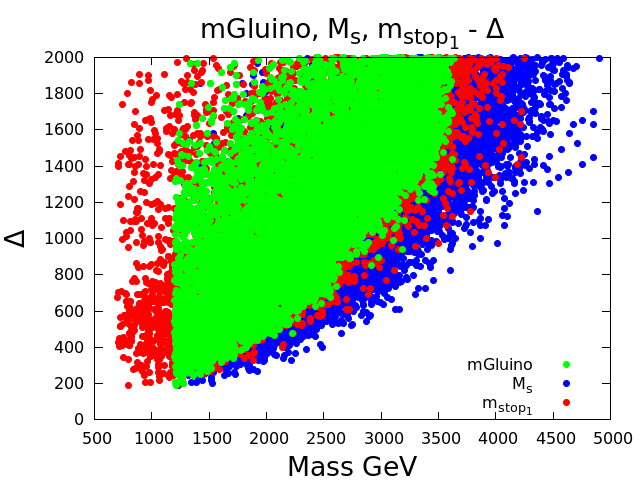}%
\end{minipage}
\end{center}
\caption{Fine tuning as a function of $m_{\tilde{g}}, M_S,$ and $m_{\text{stop}_1}$
in the NMSSM+ when $m_{h_2}$ is SM-like.}
\label{fig:nmssmp8}
\end{figure}

Furthermore, when $m_{h_2}$ varies between 123 GeV and 127 GeV, the fine tuning is a plateau around 200.
However, Figure~\ref{fig:nmssmp8} shows that the lowest fine tuning increases from $\sim 200$ to
2000 when $M_S$, $m_{\tilde{g}}$, and $m_{\tilde{t}_1}$ change from
1.2 TeV to 3.6 TeV, 1.2 TeV to 4.9 TeV, and 700 GeV to 4 TeV, respectively. 
Notice that the lightest stop can be as large as 4 TeV 
and still results in the same degree of the lowest fine tuning as that associated 
with a gluino mass of 3.6 TeV. Therefore, we again see, as expected,
the important effect the gluino has on the lowest fine tuning in the NMSSM+.
Indeed, the curves that the data points form in conjunction with the lowest fine tuning
clearly show that the gluino mass is now most relevant to the fine tuning and in fact controls it,
as opposed to the situation in the NMSSM in~\ref{ssc:nmssm}.

\subsection{NMSSM++} \label{ssc:nmssmpp}
Finally, the fine tuning in the parameter space specified in Section~\ref{sc:frame}
for the NMSSM++ is shown for the first time. The effect of having to start with 
a very large $M_3(M_{\text{GUT}})=m_{1/2}$ as explained in Equation~\ref{eq:M3} is very 
significant here in comparison with the previous two models. Particularly, the minimum mass scale of the
stops in the NMSSM++ will be larger than that in the NMSSM+.

\subsubsection{Case 1: $m_{h_1}$ is SM-like.} \label{sssc:case1pp}

The parameter space of the NMSSM++ is significantly different from
both the parameter spaces of the NMSSM and the NMSSM+. It is charactarized
by large values of $m_0$ and $m_{1/2}$ in order to be compatible with our
phenomenology constraints.  

\begin{figure}[H]
\begin{minipage}[t][1\totalheight][c]{0.45\columnwidth}%
\includegraphics[scale=0.3]{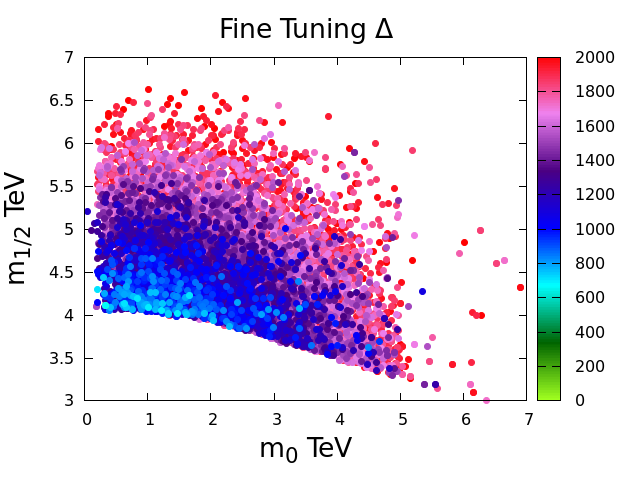}%
\end{minipage}\hfill{}%
\begin{minipage}[t][1\totalheight][c]{0.45\columnwidth}%
\includegraphics[scale=0.3]{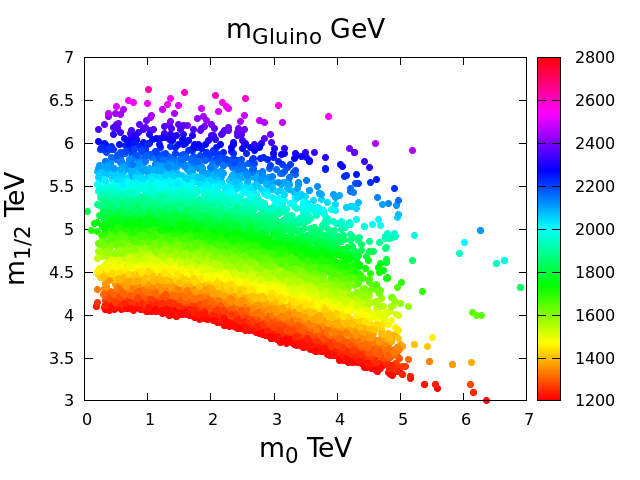}%
\end{minipage}
\caption{Left panel shows the fine tuning while the right panel shows the gluino mass, both
in the $m_0-m_{1/2}$ plane in the NMSSM++ when $m_{h_1}$ is SM-like.}
\label{fig:nmssmpp1}
\end{figure}

The fine tuning starts at a value of $\mathcal{O}(600)$, shown in Figure~\ref{fig:nmssmpp1},
and rapidly increases as $m_0$ and $m_{1/2}$ increase. In this parameter space,
the lowest fine tuning found is $\Delta \sim 663$ 
for: $m_{h_1} = 126$ GeV, $m_{\tilde{g}} = 1.2$ TeV, $m_{\tilde{t}_1} = 2.1$ TeV.

\begin{figure}[H]
\begin{minipage}[t][1\totalheight][c]{0.45\columnwidth}%
\includegraphics[scale=0.3]{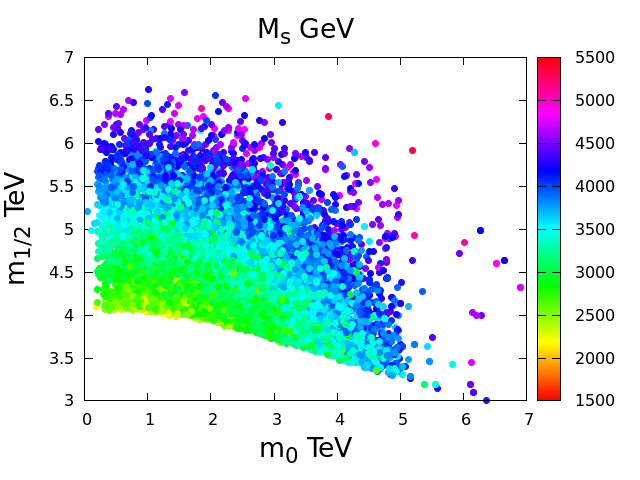}%
\end{minipage}\hfill{}%
\begin{minipage}[t][1\totalheight][c]{0.45\columnwidth}%
\includegraphics[scale=0.3]{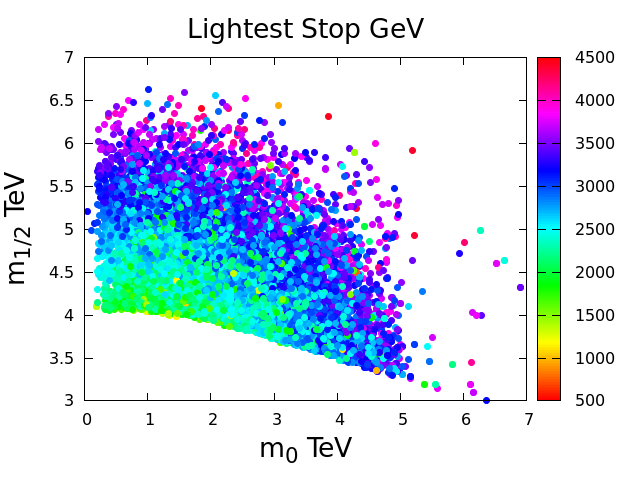}%
\end{minipage}
\caption{Left panel shows the RMS stop mass, while the right panel shows the 
lightest stop mass, both in the $m_0-m_{1/2}$ plane in the NMSSM++ when $m_{h_1}$ is SM-like.}
\label{fig:nmssmpp2}
\end{figure}

Note that a large value of $m_{1/2}$, $\sim 4$ TeV is needed to
obtain a gluino mass of 1.2 TeV. And very roughly the correlation
between $m_{1/2}$ and $m_{\tilde{g}}$ is on average: $m_{\tilde{g}} \sim 0.25 m_{1/2}$. 
Only when $m_0$ is significantly large, one can
access slightly smaller values of $m_{1/2}$. Moreover, since $m_{1/2}$
is very large it controls the scalar masses as demonstrated 
in the left panel of Figure~\ref{fig:nmssmpp2}
which shows that the parameter $M_S$ is always larger than 1.8 TeV in this parameter space,
and rises rapidly with $m_{1/2}$ to values close to $\sim 5$ TeV.

Furthermore, the mass of the lightest stop (Figure~\ref{fig:nmssmpp2}, right panel)
assumes values between 700 GeV and 4.5 TeV. However, those points with 
$m_{\tilde{t}_1} \sim 700$ GeV correspond to $m_{\tilde{t}_2} > 2.6$ TeV, and $m_{H_u}(M_{\text{GUT}}) \sim 5$ TeV.
Next, the fine tuning is almost a plateau around $600$ with respect to $m_{h_1}$. Again,
the mass of the Higgs plays no role in controlling the lowest fine tuning in the NMSSM++. 

\begin{figure}[H]
\begin{center}
\begin{minipage}[t][1\totalheight][c]{0.45\columnwidth}%
\includegraphics[scale=0.3]{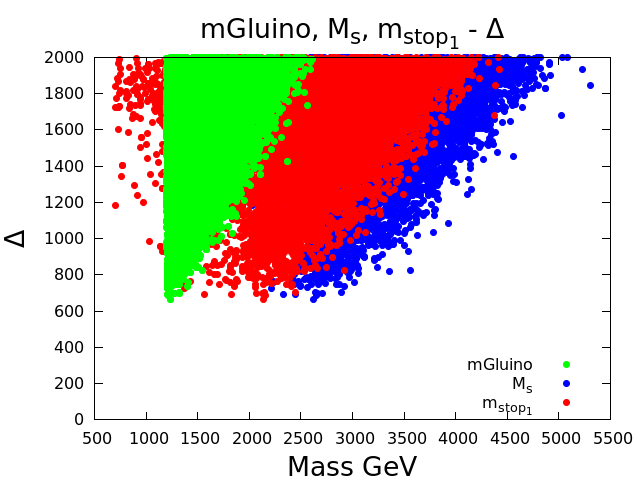}%
\end{minipage}
\end{center}
\caption{Fine tuning as a function of $m_{\tilde{g}}, M_S,$ and $m_{\text{stop}_1}$
in the NMSSM++ when $m_{h_1}$ is SM-like.}
\label{fig:nmssmpp3}
\end{figure}

On the other hand, the lowest fine tuning sharply
increases from $\sim 600$ to 2000 as $M_S$, $m_{\tilde{g}}$, and $m_{\tilde{t}_1}$ increase from 2.5 TeV
to around 4.8 TeV, 1.2 TeV to 2.6 TeV, and 2.5 TeV to 4.2 TeV, respectively as Figure~\ref{fig:nmssmpp3}
shows. Clearly, the gluino mass in the NMSSM++ strongly
drives the lowest fine tuning to be larger than that in the NMSSM and the
NMSSM+ because it raises $M_S$ to quite large values. Therefore, even though
the original goal of increasing $\lambda$ at the low scale can be easily achieved
in the NMSSM++, it comes at the expense of having very large $M_3(M_{\text{GUT}}) = m_{1/2}$
in order to obtain the gluino mass around 1.2 TeV.
Consequently, this will dominate the running of the stops, thereby making them much heavier
than the current experimental limits. This effect is the reason why the NMSSM++ (similarly the NMSSM+)
is more fine tuned than the NMSSM.
  
\subsubsection{Case 2: $m_{h_2}$ is SM-like.} \label{sssc:case2pp}

\begin{figure}[H]
\begin{minipage}[t][1\totalheight][c]{0.45\columnwidth}%
\includegraphics[scale=0.3]{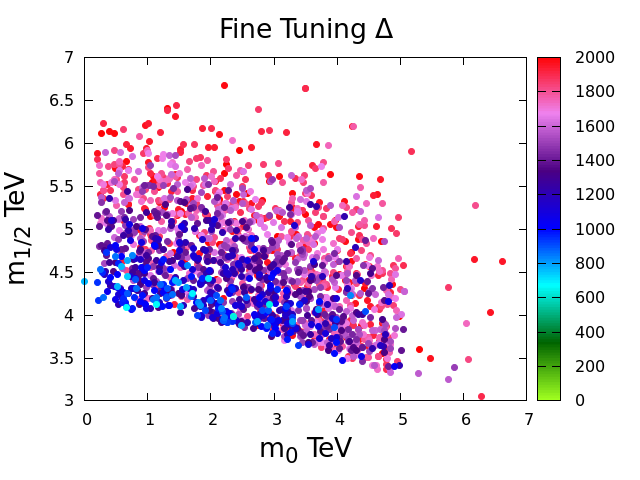}%
\end{minipage}\hfill{}%
\begin{minipage}[t][1\totalheight][c]{0.45\columnwidth}%
\includegraphics[scale=0.3]{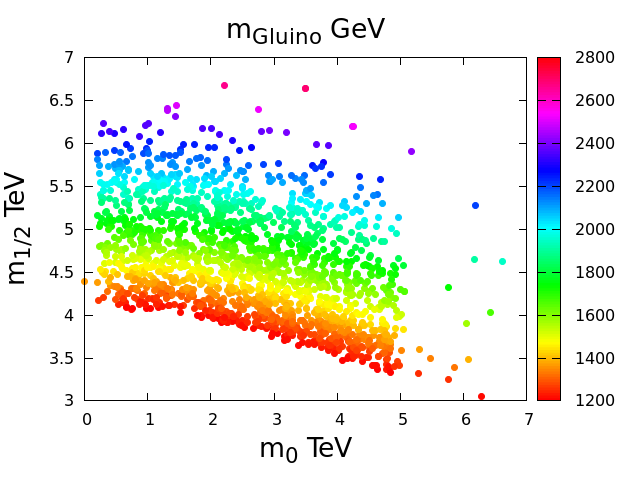}%
\end{minipage}
\caption{Left panel shows the fine tuning while the right panel shows the gluino mass, both
in the $m_0-m_{1/2}$ plane in the NMSSM++ when $m_{h_2}$ is SM-like.}
\label{fig:nmssmpp6}
\end{figure}

Here, we present the results of Case 2 where $m_{h_2}$ is SM-like in the NMSSM++.
First, due to our sampling procedure, the parameter space contains fewer points satisfying
the applied cuts than in the previous case. The fine tuning results in the
$m_0-m_{1/2}$ plane are presented in the left panel of Figure~\ref{fig:nmssmpp6}.
Overall, the patterns are similar to those found in the Case 1 in~\ref{sssc:case1pp}.
While the lowest fine tuning possible is still around 600, most of the points in this parameter
space has fine tuning above 800. The fine tuning, again, is more sensitive to changes
in $m_{1/2}$ than in $m_0$.  
The lowest fine tuning found in this parameter space is $\Delta \sim 634$ 
for: $m_{h_2} = 126$ GeV, $m_{\tilde{g}} = 1.2$ TeV, and $m_{\tilde{t}_1} = 2.76$ TeV.

The gluino mass distribution in Figure~\ref{fig:nmssmpp6} (right panel)
shows that it ranges from 1.2 TeV to 2.8 TeV. Again, very roughly and on average
$m_{\tilde{g}}\sim 0.25 m_{1/2}$. The reason this correlation is
very rough in the NMSSM++ is that we are presenting regions where $m_0$
is very large. This means that the corrections to the gluino mass due from
scalars is significant.
  
\begin{figure}[H]
\begin{minipage}[t][1\totalheight][c]{0.45\columnwidth}%
\includegraphics[scale=0.3]{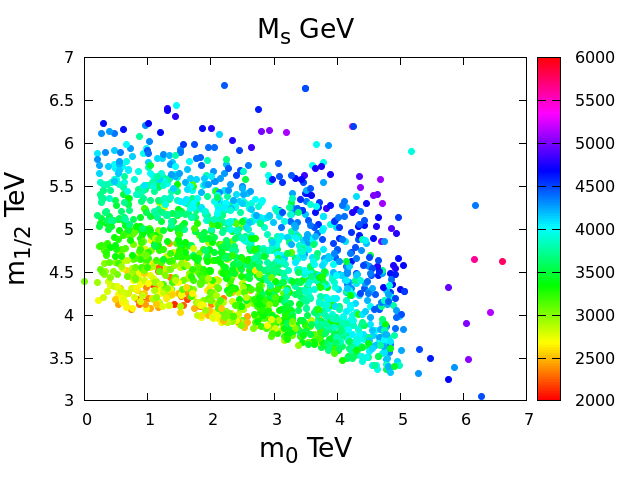}%
\end{minipage}\hfill{}%
\begin{minipage}[t][1\totalheight][c]{0.45\columnwidth}%
\includegraphics[scale=0.3]{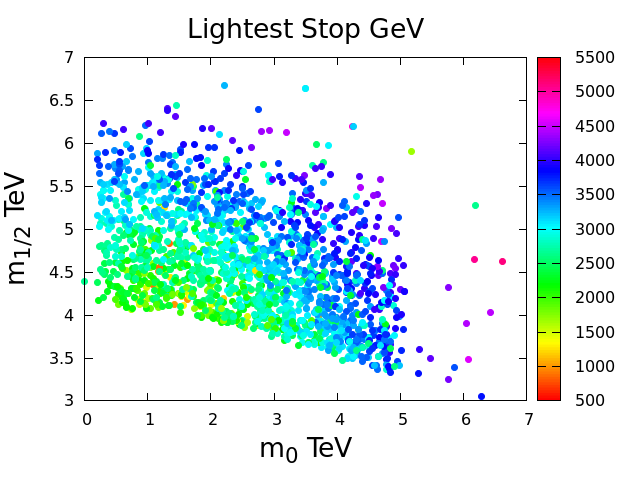}%
\end{minipage}
\caption{Left panel shows the RMS stop mass, while the right panel shows 
the lightest stop mass, both in the $m_0-m_{1/2}$ plane in the NMSSM++ when $m_{h_2}$ is SM-like.}
\label{fig:nmssmpp7}
\end{figure}

Next, the RMS stop mass,$M_S$,
see Figure~\ref{fig:nmssmpp7}, is quite
large as it starts from 2 TeV (as opposed to $900$ GeV and $1.2$ TeV in the NMSSM and the NMSSM+). 
Thus, both stops are pushed to heavy
values. Again, this is because $m_{1/2}$ has to be very large $\sim 4$ TeV in order to
satisfy the gluino mass limit. 

\begin{figure}[H]
\begin{center}
\begin{minipage}[t][1\totalheight][c]{0.45\columnwidth}%
\includegraphics[scale=0.3]{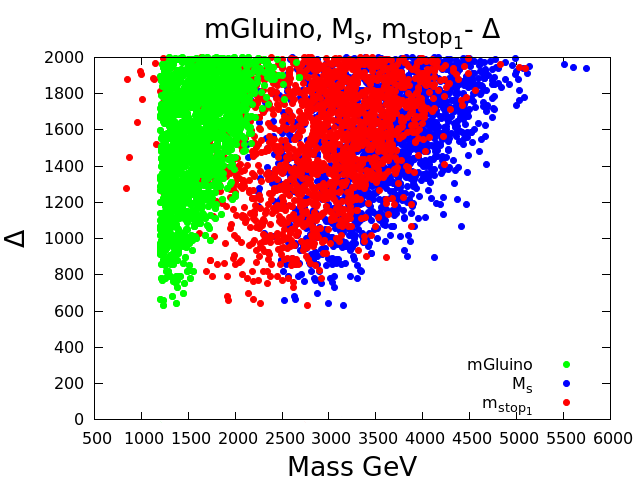}%
\end{minipage}
\end{center}
\caption{Fine tuning as a function of $m_{\tilde{g}}, M_S,$ and $m_{\text{stop}_1}$
in the NMSSM++ when $m_{h_2}$ is SM-like.}
\label{fig:nmssmpp8}
\end{figure}

Moreover, the fine tuning does not vary significantly with $m_{h_2}$ as it
is found to be $\sim 600$ for $ 123 \leq m_{h_2} \leq 127$ GeV. 
On the other hand, Figure~\ref{fig:nmssmpp8} shows that that increasing
the lightest stop from around 2 TeV to 4.5 TeV, and increasing $M_S$ from
2.5 TeV to 5 TeV, results in a raise in the fine tuning from around 600 to 2000.
More noticeably, the fine tuning increases sharply 
from around 600 to 2000 as $m_{\tilde{g}}$ increases from
1.2 TeV to around 2.8 TeV. This is a key feature of the NMSSM++ and the reason
why it is much more fine tuned than the NMSSM, and the NMSSM+. 

\subsection{Comparison} \label{ssc:comp}

\begin{figure}[H]
\begin{minipage}[t][1\totalheight][c]{0.45\columnwidth}%
\includegraphics[scale=0.3]{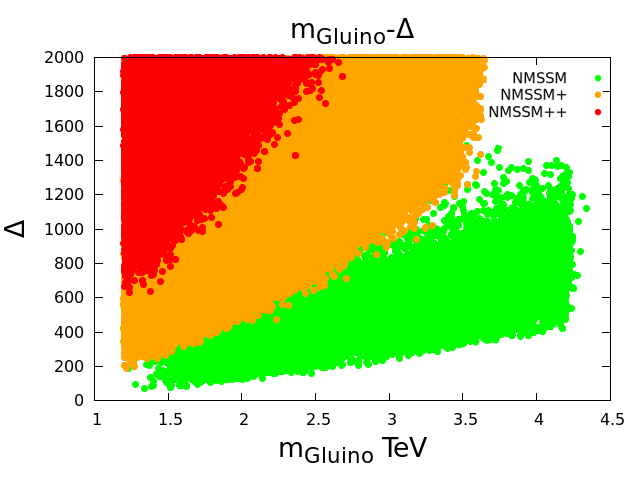}%
\end{minipage}\hfill{}%
\begin{minipage}[t][1\totalheight][c]{0.45\columnwidth}%
\includegraphics[scale=0.3]{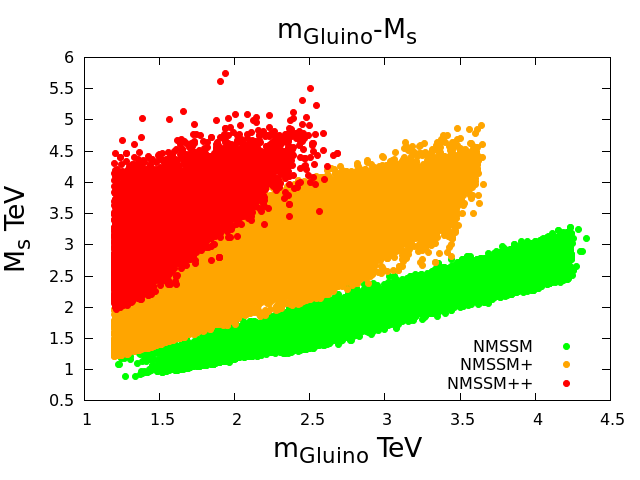}%
\end{minipage}
\caption{The left panel shows the correlation between the fine tuning and the gluino mass
for the three models. The right panel shows the correlation between the RMS stop mass and the gluino mass
for the three models.}
\label{fig:compare1}
\end{figure}

Here, we compare the three models to point out the main finding which is that adding extra matter 
to the NMSSM, hence increasing $\lambda(M_{\text{SUSY}})$, does not necessarily improve the fine tuning.
In fact, it makes it worse, especially in the framework we have 
chosen. We found that the RG running of the $\alpha_s$ and similarly the gluino 
forces one to start with a large $m_{1/2}$ ($M_3(M_{\text{GUT}})$) at the GUT scale in the plus-type
models in order to reach the desired gluino mass at the low scale. This, in turn, 
causes an increase in the mass of the stops at the low scales in comparison to 
the NMSSM as Figure~\ref{fig:compare1} shows. It is clear from this Figure that,
in all of the parameter spaces we studied, and for a given physical gluino mass, 
it is always possible to find $M_S$ that is smaller in the NMSSM than in both the NMSSM+ 
and the NMSSM++, and smaller in the NMSSM+ than in the NMSSM++. This is an RGE effect 
that was explained in Section \ref{sc:1lp}. The larger $M_S$ is, the larger the separation between 
the weak and the SUSY scales, and, as a consequence, the larger the fine tuning in 
the plus-type models, especially the NMSSM++.

The fine tuning results in the three models can be straightforwardly compared by
referring to Figures~\ref{fig:nmssm1},~\ref{fig:nmssmp1}, and~\ref{fig:nmssmpp1} for Case 1, and Figures~\ref{fig:nmssm6},
~\ref{fig:nmssmp6}, and~\ref{fig:nmssmpp6} for Case 2. Moreover, the correlation between the fine tuning  
and both of $m_{\tilde{t}_1}$ and $m_{\tilde{g}}$ in each model is shown in Figure~\ref{fig:nmssm3}, 
Figure~\ref{fig:nmssmp3}, and~\ref{fig:nmssmpp3},  for Case 1. And in
Figure~\ref{fig:nmssm8}, 
Figure~\ref{fig:nmssmp8}, and~\ref{fig:nmssmpp8}, for Case 2.

\section{Conclusions} \label{sc:con}

In this paper, we have considered three non-minimal $Z_3$-invariant supersymmetric models. Namely, the
NMSSM, the NMSSM+ that adds 3$(5,\overline{5})$ extra states of SU(5) to the NMSSM, 
and the NMSSM++ that adds 4$(5,\overline{5})$ extra states of SU(5) to the NMSSM.
Moreover, the extra matter
in the NMSSM+ and NMSSM++ is treated as a secluded sector that only affects
the mass spectrum of the ordinary sparticles through gauge interactions.
We have calculated the low energy spectrum (focusing on naturalness-related sparticles
and the SM-like Higgs boson) using the package NMSSMTools.
We have modified NMSSMTools by implementing two-loop RGEs of the NMSSM+,
and the NMSSM++. Furthermore, we have assumed that the extra matter is mass degenerate
at the scale of the first and second generations of squarks. Hence, the running masses
of the extra matter was ignored due to suppression by powers of gauge couplings and loop factors at one- and two-loop.  
    
Such extensions are known to relax the perturbativity bound on $\lambda(M_{\text{SUSY}})$,
which is the coupling between the SM singlet superfield and the up- and down-type Higgs doublets.
As a result, it is expected that the tree-level Higgs mass in the NMSSM++ will be larger than in the 
NMSSM+, and larger in the NMSSM+ than in the NMSSM without extra matter. Moreover, it is usually
assumed that the fine tuning reduces as the perturbativity bound on $\lambda$ is increased
since a large tree-level Higgs mass could imply lighter stops
in the plus-type models than in the NMSSM.
We have tested this commonly held hypothesis in the context of the three models above, and surprisingly
find that this is not the case. 
Indeed, of all three models, we find that the 
NMSSM is the least fine tuned ($\Delta \sim 100$).
The fine tuning in the NMSSM+ was the closest of the two plus-type models to the NMSSM
with the lowest value being $\Delta \sim 200$. Finally, the NMSSM++ is 
the most fine tuned model where the fine tuning starts from 600. In general,
the mass spectrum in the NMSSM++ was found to be heavier than in the NMSSM+, and heavier
in the NMSSM+ than in the NMSSM.

The reason why the fine tuning is worse in the plus-type models than in the NMSSM
is that such models with extra matter involve a larger gluino mass at high energies.
In particular, we find that $M_{3,\text{GUT}}$ is always
larger in the NMSSM+ and very much larger in the NMSSM++,
as compared to the NMSSM. This ordering results in an 
increased low energy stop mass spectrum, well above either the stop mass experimental limits or the
stop mass limits required to obtain a sufficiently large Higgs mass. The heavy stop masses
appear to be unavoidable in the NMSSM+, and especially the NMSSM++, purely as a result of the
low energy experimental gluino mass limit and the RGE running behaviour, 
at least for the class of high energy semi-constrained SUGRA inspired models under consideration.
In conclusion, it appears that increasing the perturbativity bound on $\lambda$ at the low scale
by adding extra matter does not reduce the fine tuning, but worsens it.

\section*{acknowledgements}
The work of MB is funded by King Saud University (Riyadh, Saudi Arabia). 
SFK acknowledges support from the 
European Union FP7 ITN-INVISIBLES (Marie Curie Actions, PITN- GA-2011- 289442) and the STFC Consolidated ST/J000396/1 grant.
The authors acknowledge the use of the IRIDIS High Performance Computing Facility, and associated support services at the University of Southampton, in the completion of this work.

\appendix
\section{Two-loop RGEs}
\label{A}

In this Appendix we present the two-loop RGEs (in the $\overline{\text{DR}}$ scheme) used to obtain the mass spectrum and fine tuning results.
We follow the same notation in \cite{Ellwanger:2009dp}, and use SM normalisation of the $U(1)_Y$ gauge coupling, $g_1$
in the three models. Also, $t \equiv \ln Q^2$. Finally, the RGE coefficients that are different in the three models
are placed as follows: $$\{ \text{\tiny{NMSSM}},\text{\tiny{NMSSM+}},\text{\tiny{NMSSM++}}\}$$ in the same RGE equation.
For example, the coefficients between braces in: $$16\pi^2 \frac{dg_1^2}{dt} = \underline{\{ 11,16,\frac{53}{3} \}}g_1^4 $$ 
belong to the NMSSM, NMSSM+, and NMSSM++, respectively.

Two-loop RGEs of gauge and Yukawa couplings in the NMSSM, NMSSM+ and NMSSM++ are,
\bea
16\pi^2 \frac{dg_1^2}{dt} &=& \{ 11,16,\frac{53}{3} \}g_1^4 + \frac{g_1^4}{16\pi^2}
\bigg( \{ \frac{199}{9},26,\frac{737}{27} \}g_1^2 + \{ 9,18,21 \}g_2^2 \nn \\
&+& \{ \frac{88}{3},40,\frac{392}{9} \}g_3^2 -
\frac{26}{3}h_t^2 - \frac{14}{3}h_b^2 - 6h_\tau^2 - 2\l^2 \bigg)\; , \nn
\\
16\pi^2 \frac{dg_2^2}{dt} &=&  \{1,4,5\}g_2^4 + \frac{g_2^4}{16\pi^2}
\bigg( \{3,6,7\}g_1^2 + \{25,46,53\}g_2^2 + 24g_3^2 \nn \\
&-& 6h_t^2
- 6h_b^2 - 2h_\tau^2 - 2\l^2 \bigg)\; , \nn \\
16\pi^2 \frac{dg_3^2}{dt} &=& \{-3,0,1\} g_3^4 + \frac{g_3^4}{16\pi^2} \bigg(
\{\frac{11}{3},5,\frac{49}{9}\}g_1^2 + 9g_2^2 + \{14,48,\frac{178}{3}\}g_3^2 - 4h_t^2 - 4 h_b^2 \bigg)\; , \nn
\eea
\bea
16\pi^2 \frac{dh_t^2}{dt} &=& h_t^2\bigg( 6h_t^2 + h_b^2 + \l^2
- \frac{13}{9}g_1^2 - 3g_2^2 - \frac{16}{3}g_3^2 \bigg) \nn \\
&+& \frac{h_t^2}{16\pi^2} \bigg( - 22h_t^4 - 5h_b^4 - 3\l^4 -
5h_t^2h_b^2 - 3h_t^2\l^2 - h_b^2h_\tau^2 - 4h_b^2\l^2 \nn \\
&-& h_\tau^2\l^2 - 2\l^2\k^2 +  2g_1^2h_t^2
+ \frac{2}{3}g_1^2h_b^2 + 6g_2^2h_t^2 + 16g_3^2h_t^2 \nn \\
&+& \{2743,3913,4303\} \frac{g_1^4}{162} + \{15,33,39\} \frac{g_2^4}{2} 
+ \{-16,128,176\} \frac{g_3^4}{9} \nn \\
&+& \frac{5}{3}g_1^2g_2^2 + \frac{136}{27}g_1^2g_3^2 + 8g_2^2g_3^2
\bigg)\; , \nn
\eea 
\bea
16\pi^2 \frac{dh_b^2}{dt} &=& h_b^2\bigg( 6h_b^2 + h_t^2 + h_\tau^2 +
\l^2 - \frac{7}{9}g_1^2 - 3g_2^2 - \frac{16}{3}g_3^2 \bigg) \nn \\
&+& \frac{h_b^2}{16\pi^2} \bigg( - 22h_b^4 - 5h_t^4
- 3h_\tau^4 - 3\l^4 - 5h_b^2h_t^2 - 3h_b^2h_\tau^2 - 3h_b^2\l^2 \nn \\
&-& 4h_t^2\l^2 - 2\l^2\k^2 + \frac{2}{3}g_1^2h_b^2 +
\frac{4}{3}g_1^2h_t^2 + 2g_1^2h_\tau^2 + 6g_2^2h_b^2 + 16g_3^2h_b^2 \nn
\\
&+& \{1435,2065,2275\} \frac{g_1^4}{162} + \{15,33,39\} \frac{g_2^4}{2} + \{-16,128,176\} \frac{g_3^4}{9} \nn \\
&+& \frac{5}{3}g_1^2g_2^2 + \frac{40}{27}g_1^2g_3^2 + 8g_2^2g_3^2\bigg)\;
, \nn \\
16\pi^2 \frac{dh_\tau^2}{dt} &=& h_\tau^2\Big( 4h_\tau^2 + 3h_b^2 + \l^2
- 3g_1^2 - 3g_2^2 \Big) \nn \\
&+& \frac{h_\tau^2}{16\pi^2} \bigg( - 10h_\tau^4 - 9h_b^4 - 3\l^4
- 9h_\tau^2h_b^2 - 3h_\tau^2\l^2 - 3h_t^2h_b^2 - 3h_t^2\l^2 \nn \\
&-& 2\l^2\k^2  + 2g_1^2h_\tau^2 - \frac{2}{3}g_1^2h_b^2
+ 6g_2^2h_\tau^2 + 16g_3^2h_b^2 \nn \\
&+& \{75,105,115\} \frac{g_1^4}{2}
+ \{15,33,39\}\frac{g_2^4}{2} + 3g_1^2g_2^2 \bigg)\; , \nn \\
16\pi^2 \frac{d\l^2}{dt} &=& \l^2\Big(3h_t^2 + 3h_b^2 + h_\tau^2 +4\l^2
+ 2\k^2 - g_1^2 - 3g_2^2 \Big) \nn \\
&+& \frac{\l^2}{16\pi^2} \bigg( - 10\l^4  - 9h_t^4 - 9h_b^4
- 3h_\tau^4 - 8\k^4 - 9\l^2h_t^2 - 9\l^2h_b^2 \nn \\
&-& 3\l^2h_\tau^2 -12\l^2\k^2 - 6h_t^2h_b^2 + 2g_1^2\l^2 +
\frac{4}{3}g_1^2h_t^2 - \frac{2}{3}g_1^2h_b^2 + 2g_1^2h_\tau^2 \nn \\
&+& 6g_2^2\l^2 + 16g_3^2h_t^2 + 16g_3^2h_b^2
+ \{23,33,\frac{109}{3}\}\frac{g_1^4}{2} + \{15,33,\frac{117}{3}\}\frac{g_2^4}{2} + 3g_1^2g_2^2 \bigg)\; , \nn \\
16\pi^2 \frac{d\k^2}{dt} &=& \k^2\Big(6\l^2 +6\k^2\Big)
+ \frac{\k^2}{16\pi^2} \bigg( - 24\k^4 - 12\l^4 - 24\k^2\l^2 \nn \\
&-& 18h_t^2\l^2 - 18h_b^2\l^2 - 6h_\tau^2\l^2 + 6g_1^2\l^2 + 18g_2^2\l^2
\bigg)\; . \nn
\eea
Two-loop RGEs for the gauginos in the NMSSM, NMSSM+, and NMSSM++ are,
\bea
16\pi^2 \frac{dM_1}{dt} &=& \{ 11,16,\frac{53}{3} \} g_1^2M_1
+ \frac{g_1^2}{16\pi^2} \bigg( \frac{398}{9},52,\frac{1474}{27}\}g_1^2M_1
+\{9,18,21\}g_2^2\big(M_1+M_2\big) \nn \\
&+& \{\frac{88}{3},40,\frac{392}{9}\}g_3^2\big(M_1+M_3 \big) \nn \\
&+& \frac{26}{3}h_t^2\big(A_t-M_1\big) +
\frac{14}{3}h_b^2\big(A_b-M_1\big)
+ 6h_\tau^2\big(A_\tau-M_1\big) + 2\l^2\big(A_\l-M_1\big)\bigg)\; , \nn
\\
16\pi^2 \frac{dM_2}{dt} &=& \{1,4,5\}g_2^2M_2
+ \frac{g_2^2}{16\pi^2} \bigg( \{3,6,7\}g_1^2\big(M_1+M_2\big)
+ \{50,92,106\}g_2^2M_2 \nn \\
&+& 24g_3^2\big(M_2+M_3\big)
+ 6h_t^2\big(A_t-M_2\big) + 6h_b^2\big(A_b-M_2\big) \nn \\
&+& 2h_\tau^2\big(A_\tau-M_2\big) + 2\l^2\big(A_\l-M_2\big) \bigg)\; , \nn
\\
16\pi^2 \frac{dM_3}{dt} &=& \{-3,0,1\}g_3^2M_3 
+ \frac{g_3^2}{16\pi^2} \bigg( \{\frac{11}{3},5,\frac{49}{9}\}g_1^2\big(M_1+M_3\big)
+9g_2^2\big(M_2+M_3\big) \nn \\ 
&+& \{28,96,\frac{356}{3}\}g_3^2M_3 + 4h_t^2\big(A_t-M_3\big) + 4h_b^2\big(A_b-M_3\big) \bigg)\; . \nn
\eea
Two-loop RGEs of the trilinear couplings in the NMSSM, NMSSM+, and NMSSM++ are,
\bea
16\pi^2 \frac{dA_t}{dt} &=& 6h_t^2A_t + h_b^2A_b  + \l^2A_\l
+ \frac{13}{9}g_1^2M_1 + 3g_2^2M_2 + \frac{16}{3}g_3^2M_3\nn \\
&+& \frac{1}{16\pi^2} \bigg( - 44h_t^4A_t - 10 h_b^4A_b - 6\l^4A_\l
- 5h_t^2h_b^2\big(A_t+A_b\big) \nn \\
&-& 3h_t^2\l^2\big(A_t+A_\l\big) - h_b^2h_\tau^2\big(A_b+A_\tau\big)
- 4h_b^2\l^2\big(A_b+A_\l\big) \nn \\
&-& h_\tau^2\l^2\big(A_\tau+A_\l\big) - 2\l^2\k^2\big(A_\l+A_\k\big)
+ 2g_1^2h_t^2\big(A_t-M_1\big) \nn \\
&+& \frac{2}{3}g_1^2h_b^2\big(A_b-M_1\big) + 6g_2^2h_t^2\big(A_t-M_2\big)
+ 16g_3^2h_t^2\big(A_t-M_3\big) \nn \\
&-& \{2743,3919,4303\}\frac{g_1^4M_1}{81} - \{15,33,39\}g_2^4M_2 + \{32,-256, -352\}\frac{g_3^4M_3}{9} \nn \\
&-& \frac{5}{3}g_1^2g_2^2\big(M_1+M_2\big)
- \frac{136}{27}g_1^2g_3^2\big(M_1+M_3\big)
- 8g_2^2g_3^2\big(M_2+M_3\big) \bigg)\; , \nn
\eea
\bea
16\pi^2 \frac{dA_b}{dt} &=& 6h_b^2A_b + h_t^2A_t + h_\tau^2A_\tau
+ \l^2A_\l +\frac{7}{9}g_1^2M_1 + 3g_2^2M_2 + \frac{16}{3}g_3^2M_3\nn \\
&+& \frac{1}{16\pi^2} \bigg( - 44h_b^4A_b - 10 h_t^4A_t
- 6h_\tau^4A_\tau - 6\l^4A_\l - 5h_b^2h_t^2\big(A_b+A_t\big) \nn \\
&-& 3h_b^2h_\tau^2\big(A_b+A_\tau\big) -  3h_b^2\l^2\big(A_b+A_\l\big)
- 4h_t^2\l^2\big(A_t+A_\l\big) \nn \\
&-& 2\l^2\k^2\big(A_\l+A_\k\big) + \frac{2}{3}g_1^2h_b^2\big(A_b-M_1\big)
+ \frac{4}{3}g_1^2h_t^2\big(A_t-M_1\big) \nn \\
&+& 2g_1^2h_\tau^2\big(A_\tau-M_1\big) + 6g_2^2h_b^2\big(A_b-M_2\big)
+ 16g_3^2h_b^2\big(A_b-M_3\big) \nn \\
&-& \{1435,2065,2275\}\frac{g_1^4M_1}{81} - \{15,33,39\}g_2^4M_2 + \{32,-256, -352\}\frac{g_3^4M_3}{9} \nn \\
&-& \frac{5}{3}g_1^2g_2^2\big(M_1+M_2\big)
- \frac{40}{27}g_1^2g_3^2\big(M_1+M_3\big)
- 8g_2^2g_3^2\big(M_2+M_3\big) \bigg)\; , \nn \\
16\pi^2 \frac{dA_\tau}{dt} &=& 4h_\tau^2A_\tau + 3h_b^2A_b
+ \l^2A_\l + 3g_1^2M_1 + 3g_2^2M_2 + \frac{1}{16\pi^2} \bigg( \nn \\
&-& 20h_\tau^4A_\tau - 18 h_b^4A_b - 6\l^4A_\l
- 9h_\tau^2h_b^2\big(A_\tau+A_b\big) - 3h_\tau^2\l^2\big(A_\tau+A_\l\big) \nn \\
&-& 3h_t^2h_b^2\big(A_t+A_b\big) - 3h_t^2\l^2\big(A_t+A_\l\big)
- 2\l^2\k^2\big(A_\l+A_\k\big)  \nn \\
&+& 2g_1^2h_\tau^2\big(A_\tau-M_1\big) - \frac{2}{3}g_1^2h_b^2\big(A_b-M_1\big)
+ 6g_2^2h_\tau^2\big(A_\tau-M_2\big) \nn \\
&+&  16g_3^2h_b^2\big(A_b-M_3\big) - \{75,105,115\}g_1^4M_1 \nn \\
&-& \{15,33,39\}g_2^4M_2
- 3g_1^2g_2^2\big(M_1+M_2\big) \bigg)\; , \nn \\
\eea
\bea
16\pi^2 \frac{dA_\l}{dt} &=& 4\l^2A_\l + 3h_t^2A_t + 3h_b^2A_b
+ h_\tau^2A_\tau + 2\k^2A_\k + g_1^2M_1 + 3g_2^2M_2 \nn \\
&+& \frac{1}{16\pi^2} \bigg( - 20 \l^4A_\l - 18h_t^4A_t  - 18h_b^4A_b
- 6h_\tau^4A_\tau - 16\k^4A_\k \nn \\
&-& 9\l^2h_t^2\big(A_\l+A_t\big) - 9\l^2h_b^2\big(A_\l+A_b\big)
- 3 \l^2h_\tau^2\big(A_\l+A_\tau\big) \nn \\
&-& 12\l^2\k^2\big(A_\l+A_\k\big) - 6h_t^2h_b^2\big(A_t+A_b\big)
+ 2g_1^2\l^2(A_\l-M_1) \nn \\
&+& \frac{4}{3}g_1^2h_t^2\big(A_t-M_1\big)
- \frac{2}{3}g_1^2h_b^2\big(A_b-M_1\big)
+ 2g_1^2h_\tau^2\big(A_\tau-M_1\big) \nn \\
&+& 6g_2^2\l^2\big(A_\l-M_2\big) + 16g_3^2h_t^2\big(A_t-M_3\big)
+ 16g_3^2h_b^2\big(A_b-M_3\big) \nn \\
&-& \{23,33,\frac{109}{3} \}g_1^4M_1 - \{15,33,39 \}g_2^4M_2
- 3g_1^2g_2^2\big(M_1+M_2\big) \bigg)\; , \nn \\
16\pi^2 \frac{dA_\k}{dt} &=& 6\k^2A_\k  + 6\l^2A_\l
+ \frac{1}{16\pi^2} \bigg( - 48\k^4A_\k - 24 \l^4A_\l \nn \\
&-& 24\k^2\l^2\big(A_\k+A_\l\big)
- 18h_t^2\l^2\big(A_t+A_\l\big) - 18h_b^2\l^2\big(A_b+A_\l\big) \nn \\
&-& 6h_\tau^2\l^2\big(A_\tau+A_\l\big)
+ 6g_1^2\l^2\big(A_\l-M_1\big) + 18g_2^2\l^2\big(A_\l-M_2\big) \bigg)\;
.
\eea

Again, we follow the notation in \cite{Ellwanger:2009dp} defining: 
\bea \label{app:a3}
M_t^2 &= & m_{Q_3}^2+m_{U_3}^2+m_{H_u}^2+A_t^2\; , \nn \\
M_b^2 &= & m_{Q_3}^2+m_{D_3}^2+m_{H_d}^2+A_b^2\; , \nn \\
M_\tau^2 &= & m_{L_3}^2+m_{E_3}^2+m_{H_d}^2+A_\tau^2\; , \nn \\
M_\l^2 &= & m_{H_u}^2+m_{H_d}^2+m_S^2+A_\l^2\; , \nn \\
M_\k^2 &= & 3m_S^2+A_\k^2\; , \nn \\
\xi &=& {\rm Tr}\big[{\bf m}_Q^2 - 2{\bf m}_U^2 + {\bf m}_D^2
- {\bf m}_L^2 + {\bf m}_E^2 + \{  {\bf m}_{\bar{Dx}}^2 -{\bf m}_{Dx}^2 + {\bf m}_{H_{ux}}^2 - {\bf m}_{H_{dx}}^2 \} \big]
+ m_{H_u}^2 - m_{H_d}^2\; , \nn \\
\xi' &=& h_t^2\big(-m_{Q_3}^2+4m_{U_3}^2-3m_{H_u}^2\big)
+ h_b^2\big(-m_{Q_3}^2-2m_{D_3}^2+3m_{H_d}^2\big) \nn \\
&+& h_\tau^2\big(m_{L_3}^2-2m_{E_3}^2+m_{H_d}^2\big)
+ \l^2\big(m_{H_d}^2-m_{H_u}^2\big) \nn \\
&+& g_1^2 \Big( {\rm Tr}\Big[\frac{1}{18}{\bf m}_Q^2
- \frac{16}{9}{\bf m}_U^2 + \frac{2}{9}({\bf m}_D^2 + \{{\bf m}_{\bar{Dx}}^2 - {\bf m}_{Dx}^2 \}) - \frac{1}{2}({\bf
m}_L^2 + \{+ {\bf m}_{H_{ux}}^2 - {\bf m}_{H_{dx}}^2 \}) + 2{\bf m}_E^2\Big] \nn \\
&+& \frac{1}{2}\big(m_{H_u}^2 - m_{H_d}^2\big)
\Big) \nn \\
&+& \frac{3}{2}g_2^2 \Big( {\rm Tr}\big[{\bf m}_Q^2 - {\bf m}_L^2 + \{ {\bf m}_{H_{ux}}^2 - {\bf m}_{H_{dx}}^2 \} \big]
+ m_{H_u}^2 - m_{H_d}^2 \Big) \nn \\
&+& \frac{8}{3}g_3^2{\rm Tr}\big[ {\bf
m}_Q^2 - 2{\bf m}_U^2 + {\bf m}_D^2 \{ + {\bf m}_{\bar{Dx}}^2 -{\bf m}_{Dx}^2  \} \big]\; , \nn \\
\sigma_1 &=& g_1^2 \Big( {\rm Tr}\Big[\frac{1}{3}{\bf m}_Q^2
+ \frac{8}{3}{\bf m}_U^2 + \frac{2}{3}({\bf m}_D^2 +\{{\bf m}_{Dx}^2 + {\bf m}_{\bar{Dx}}^2 \} ) \nn \\
&+& {\bf m}_L^2 + 2{\bf
m}_E^2 + \{ {\bf m}_{H_{ux}}^2 + {\bf m}_{H_{dx}}^2 \} \Big]
+ m_{H_u}^2 + m_{H_d}^2 \Big)\; , \nn \\
\sigma_2 &=& g_2^2 \Big( {\rm Tr}\big[3{\bf m}_Q^2 + {\bf m}_L^2 + \{ {\bf m}_{H_{ux}}^2 + {\bf m}_{H_{dx}}^2 \}  \big]
+ m_{H_u}^2 + m_{H_d}^2 \Big)\; , \nn \\
\sigma_3 &=& g_3^2 {\rm Tr}\big[2{\bf m}_Q^2 + {\bf m}_U^2 + {\bf
m}_D^2 + \{{\bf m}_{Dx}^2 + {\bf m}_{\bar{Dx}}^2 \}\big]\; .
\eea
\noindent where $\{{\bf m}_{\bar{Dx}}^2, {\bf m}_{Dx}^2, {\bf m}_{H_{ux}}^2, {\bf m}_{H_{dx}}^2  \}$ are
diagonal $3 \times 3$ matrices in the NMSSM+, and diagonal $4 \times 4$ matrices in the NMSSM++. 

The two-loop RGEs of the scalars in the NMSSM, NMSSM+, and NMSSM++ are,
\bea
16\pi^2 \frac{dm_{Q_a}^2}{dt} &=&
\delta_{a3}h_t^2M_t^2 + \delta_{a3}h_b^2M_b^2 - \frac{1}{9}g_1^2M_1^2
- 3g_2^2M_2^2 - \frac{16}{3}g_3^2M_3^2 + \frac{1}{6}g_1^2\xi \nn \\
&+& \frac{1}{16\pi^2} \bigg( - 10\delta_{a3}h_t^4\big(M_t^2+A_t^2\big)
- 10\delta_{a3}h_b^4\big(M_b^2+A_b^2\big) \nn \\
&-& \delta_{a3}h_t^2\l^2\big(M_t^2+M_\l^2+2A_tA_\l\big)
- \delta_{a3}h_b^2h_\tau^2\big(M_b^2+M_\tau^2+2A_bA_\tau\big) \nn \\
&-& \delta_{a3}h_b^2\l^2\big(M_b^2+M_\l^2+2A_bA_\l\big)
+ \frac{4}{3}\delta_{a3}g_1^2h_t^2\big(M_t^2-2M_1(A_t-M_1)\big) \nn \\
&+& \frac{2}{3}\delta_{a3}g_1^2h_b^2\big(M_b^2-2M_1(A_b-M_1)\big) \nn \\
&+& \{199,289,319 \} \frac{g_1^4M_1^2}{54} + \{33,87,105\} \frac{g_2^4M_2^2}{2}
+ \{-64,80,128 \} \frac{g_3^4M_3^2}{3} \nn \\
&+& \frac{1}{3}g_1^2g_2^2(M_1^2+M_2^2+M_1M_2)
+ \frac{16}{27}g_1^2g_3^2(M_1^2+M_3^2+M_1M_3) \nn \\
&+& 16g_2^2g_3^2(M_2^2+M_3^2+M_2M_3)
+ \frac{1}{3}g_1^2\xi' + \frac{1}{18}g_1^2\sigma_1
+ \frac{3}{2}g_2^2\sigma_2 + \frac{8}{3}g_3^2\sigma_3 
\bigg)\; , \nn
\eea
\bea
16\pi^2 \frac{dm_{U_a}^2}{dt} &=&
2\delta_{a3}h_t^2M_t^2 - \frac{16}{9}g_1^2M_1^2
- \frac{16}{3}g_3^2M_3^2 - \frac{2}{3}g_1^2\xi \nn \\
&+& \frac{1}{16\pi^2} \bigg( - 16\delta_{a3}h_t^4\big(M_t^2+A_t^2\big)
- 2\delta_{a3}h_t^2h_b^2\big(M_t^2+M_b^2+2A_tA_b\big) \nn \\
&-& 2\delta_{a3}h_t^2\l^2\big(M_t^2+M_\l^2+2A_tA_\l\big)
- \frac{2}{3}\delta_{a3}g_1^2h_t^2\big(M_t^2-2M_1(A_t-M_1)\big) \nn \\
&+& 6 \delta_{a3}g_2^2h_t^2\big(M_t^2-2M_2(A_t-M_2)\big) \nn \\
&+& \{1712, 2432,2672\}\frac{g_1^4M_1^2}{27} + \{-64,80,128\} \frac{g_3^4M_3^2}{3} \nn \\
&+& \frac{256}{27}g_1^2g_3^2(M_1^2+M_3^2+M_1M_3)
- \frac{4}{3}g_1^2\xi' + \frac{8}{9}g_1^2\sigma_1
+ \frac{8}{3}g_3^2\sigma_3
\bigg)\; , \nn \\
16\pi^2 \frac{dm_{D_a}^2}{dt} &=&
2\delta_{a3}h_b^2M_b^2 - \frac{4}{9}g_1^2M_1^2
- \frac{16}{3}g_3^2M_3^2 + \frac{1}{3}g_1^2\xi \nn \\
&+& \frac{1}{16\pi^2} \bigg( - 16\delta_{a3}h_b^4\big(M_b^2+A_b^2\big)
- 2\delta_{a3}h_b^2h_t^2\big(M_b^2+M_t^2+2A_bA_t\big) \nn \\
&-& 2\delta_{a3}h_b^2h_\tau^2\big(M_b^2+M_\tau^2+2A_bA_\tau\big)
- 2\delta_{a3}h_b^2\l^2\big(M_b^2+M_\l^2+2A_bA_\l\big) \nn \\
&+& \frac{2}{3}\delta_{a3}g_1^2h_b^2\big(M_b^2-2M_1(A_b-M_1)\big)
+ 6 \delta_{a3}g_2^2h_b^2\big(M_b^2-2M_2(A_b-M_2)\big) \nn \\
&+& \{404,584,644\}\frac{g_1^4M_1^2}{27} + \{-64,80,128\} \frac{g_3^4M_3^2}{3} \nn \\
&+& \frac{64}{27}g_1^2g_3^2(M_1^2+M_3^2+M_1M_3) \nn \\
&+& \frac{2}{3}g_1^2\xi' + \frac{2}{9}g_1^2\sigma_1
+ \frac{8}{3}g_3^2\sigma_3
\bigg)\; , \nn \\
16\pi^2 \frac{dm_{L_a}^2}{dt} &=&
\delta_{a3}h_\tau^2M_\tau^2 - g_1^2M_1^2 - 3g_2^2M_2^2 - \frac{1}{2}g_1^2\xi
+ \frac{1}{16\pi^2} \bigg( - 6\delta_{a3}h_\tau^4\big(M_\tau^2+A_\tau^2\big) \nn \\
&-& 3\delta_{a3}h_\tau^2h_b^2\big(M_\tau^2+M_b^2+2A_\tau A_b\big)
- \delta_{a3}h_\tau^2\l^2\big(M_\tau^2+M_\l^2+2A_\tau A_\l\big) \nn \\
&+& 2\delta_{a3}g_1^2h_\tau^2\big(M_\tau^2-2M_1(A_\tau-M_1)\big)
+ \{69,99,109\}\frac{g_1^4M_1^2}{2} + \{33,87,105\}\frac{g_2^4M_2^2}{2} \nn \\
&+& 3g_1^2g_2^2(M_1^2+M_2^2+M_1M_2)
- g_1^2\xi' + \frac{1}{2}g_1^2\sigma_1
+ \frac{3}{2}g_2^2\sigma_2
\bigg)\; , \nn
\eea
\bea
16\pi^2 \frac{dm_{E_a}^2}{dt} &=&
2\delta_{a3}h_\tau^2M_\tau^2 - 4g_1^2M_1^2 + g_1^2\xi
+ \frac{1}{16\pi^2} \bigg( - 8\delta_{a3}h_\tau^4\big(M_\tau^2+A_\tau^2\big) \nn \\
&-& 6\delta_{a3}h_\tau^2h_b^2\big(M_\tau^2+M_b^2+2A_\tau A_b\big)
- 2\delta_{a3}h_\tau^2\l^2\big(M_\tau^2+M_\l^2+2A_\tau A_\l\big) \nn \\
&-& 2\delta_{a3}g_1^2h_\tau^2\big(M_\tau^2-2M_1(A_\tau-M_1)\big)
+ 6\delta_{a3}g_2^2h_\tau^2\big(M_\tau^2-2M_2(A_\tau-M_2)\big) \nn \\
&+& \{156,216,236\}g_1^4M_1^2 + 2g_1^2\xi' + 2g_1^2\sigma_1
\bigg)\; .
\eea
The Two-loop RGEs of the Higgs doublets and singlet in the NMSSM, NMSSM+, and NMSSM++ are,
\bea
16\pi^2 \frac{dm_{H_u}^2}{dt} &=&
3h_t^2M_t^2 + \l^2M_\l^2 - g_1^2M_1^2 - 3g_2^2M_2^2 + \frac{1}{2}g_1^2\xi \nn \\
&+& \frac{1}{16\pi^2} \bigg( - 18h_t^4\big(M_t^2+A_t^2\big)
- 6\l^4\big(M_\l^2+A_\l^2\big) \nn \\
&-& 3h_t^2h_b^2\big(M_t^2+M_b^2+2A_tA_b\big)
- 3h_b^2\l^2\big(M_b^2+M_\l^2+2A_bA_\l\big) \nn \\
&-& h_\tau^2\l^2\big(M_\tau^2+M_\l^2+2A_\tau A_\l\big)
-  2\l^2\k^2\big(M_\l^2+M_\k^2+2A_\l A_\k\big) \nn \\
&+& \frac{4}{3}g_1^2h_t^2\big(M_t^2-2M_1(A_t-M_1)\big)
+ 16g_3^2h_t^2\big(M_t^2-2M_3(A_t-M_3)\big) \nn \\
&+& \{69,99,109\}\frac{g_1^4M_1^2}{2} + \{33,87,105\}\frac{g_2^4M_2^2}{2} \nn \\
&+& 3g_1^2g_2^2(M_1^2+M_2^2+M_1M_2) \nn \\
&+& g_1^2\xi' + \frac{1}{2}g_1^2\sigma_1
+ \frac{3}{2}g_2^2\sigma_2
\bigg)\; , \nn \\
16\pi^2 \frac{dm_{H_d}^2}{dt} &=&
3h_b^2M_b^2 + h_\tau^2M_\tau^2 + \l^2M_\l^2 - g_1^2M_1^2
- 3g_2^2M_2^2 - \frac{1}{2}g_1^2\xi \nn \\
&+& \frac{1}{16\pi^2} \bigg( - 18h_b^4\big(M_b^2+A_b^2\big)
- 6h_\tau^4\big(M_\tau^2+A_\tau^2\big) \nn \\
&-& 6\l^4\big(M_\l^2+A_\l^2\big)
- 3h_b^2h_t^2\big(M_b^2+M_t^2+2A_bA_t\big) \nn \\
&-& 3h_t^2\l^2\big(M_t^2+M_\l^2+2A_tA_\l\big)
-  2\l^2\k^2\big(M_\l^2+M_\k^2+2A_\l A_\k\big) \nn \\
&-& \frac{2}{3}g_1^2h_b^2\big(M_b^2-2M_1(A_b-M_1)\big)
+ 2g_1^2h_\tau^2\big(M_\tau^2-2M_1(A_\tau-M_1)\big) \nn \\
&+& 16g_3^2h_b^2\big(M_b^2-2M_3(A_b-M_3)\big) \nn \\
&+&\{69,99,109\} \frac{g_1^4M_1^2}{2} + \{33,87,105\}\frac{g_2^4M_2^2}{2} \nn \\
&+& 3g_1^2g_2^2(M_1^2+M_2^2+M_1M_2)
- g_1^2\xi' + \frac{1}{2}g_1^2\sigma_1
+ \frac{3}{2}g_2^2\sigma_2
\bigg)\; , \nn \\
16\pi^2 \frac{dm_S^2}{dt} &=&
2\l^2M_\l^2 + 2\k^2M_\k^2
+ \frac{1}{16\pi^2} \bigg( - 8\l^4\big(M_\l^2+A_\l^2\big)
- 16\k^4\big(M_\k^2+A_\k^2\big) \nn \\
&-& 6\l^2h_t^2\big(M_\l^2+M_t^2+2A_\l A_t\big)
- 6\l^2h_b^2\big(M_\l^2+M_b^2+2A_\l A_b\big) \nn \\
&-& 2\l^2h_\tau^2\big(M_\l^2+M_\tau^2+2A_\l A_\tau\big)
- 8\l^2\k^2\big(M_\l^2+M_\k^2+2A_\l A_\k\big) \nn \\
&+& 2g_1^2\l^2\big(M_\l^2-2M_1(A_\l-M_1)\big)
+ 6g_2^2\l^2\big(M_\l^2-2M_2(A_\l-M_2)\big)
\bigg)\; .\
\eea

Finally, for completeness we add the one-loop RGEs of the extra matter, in the NMSSM+ and the NMSSM++
\bea
16\pi^2 \frac{dm_{H_{ux}^i}^2}{dt} &=&
- g_1^2M_1^2 - 3g_2^2M_2^2 + \frac{1}{2}g_1^2\xi \; , \nn \\
16\pi^2 \frac{dm_{H_{dx}^i}^2}{dt} &=&
- g_1^2M_1^2 - 3g_2^2M_2^2 - \frac{1}{2}g_1^2\xi \; , \nn \\
16\pi^2 \frac{dm_{Dx^i}^2}{dt} &=&
- \frac{4}{9}g_1^2M_1^2 - \frac{16}{3}g_3^2M_3^2 - \frac{1}{3}g_1^2\xi \; , \nn \\
16\pi^2 \frac{dm_{\bar{Dx}^i}^2}{dt} &=&
- \frac{4}{9}g_1^2M_1^2 - \frac{16}{3}g_3^2M_3^2 + \frac{1}{3}g_1^2\xi \; .\
\eea

\bibliographystyle{apsrev4-1}
\bibliography{refs}

\end{document}